\documentclass[rmp,aps,preprint,nofootinbib,endfloats]{revtex4}

\usepackage{graphics}
\usepackage{epsfig}

\def\inbar{\,\vrule height1.5ex width.4pt depth0pt}
\def\IR{\relax{\rm I\kern-.18em R}}
\def\IC{\relax\hbox{$\inbar\kern-.3em{\rm C}$}}

\begin{document}
\title{A practical guide to Basic Statistical Techniques for Data Analysis in Cosmology}

\author{Licia Verde}
\email{verde@ieec.uab.es}
\affiliation{ICREA \& Institute of Space Sciences (IEEC-CSIC), \\ 
Fac. Ciencies,  Campus UAB Torre C5 parell 2 Bellaterra (Spain)}

\begin{abstract}  
This is the summary of 4 Lectures given at the XIX Canary islands winter school  of Astrophysics "The  Cosmic Microwave Background, from quantum Fluctuations  to the present Universe".
\end{abstract}                                                                 

\maketitle
\tableofcontents

\section{INTRODUCTION}
\label{sec:intro}
Statistics is everywhere in Cosmology, today more than ever:  cosmological data sets  are getting ever larger,  data from different  experiments can be compared and  combined; as the statistical error bars shrink, the effect of systematics need to be described, quantified and accounted for.  As the data sets improve the parameter space we want to explore also grows. In the last 5 years there were more than 370 papers with ''statistic-" in the title!

There are many excellent books on statistics: however when I started using statistics in cosmology I could not find all the information I needed in the same place. So here I have tried to put together a "starter kit" of statistical tools useful for cosmology.
This will not be a rigorous  introduction with theorems, proofs etc. The goal of these lectures is to {\it a)} be a practical manual: to give you enough knowledge to be able to understand cosmological data analysis and/or  to find out more by yourself and {\it b)} give you a ''bag of tricks" (hopefully) useful for your future work. 

Many useful applications are left as exercises (often with hints) and are  thus proposed in the main text. 

 I will start by introducing probability and statistics from a Cosmologist point of view. Then I will continue with the description of random fields (ubiquitous in Cosmology), followed by an introduction to Monte Carlo methods including Monte-Carlo error estimates and Monte Carlo Markov Chains. I will conclude with the Fisher matrix  technique, useful for quickly forecasting the performance of future experiments.

\section{Probabilities}

\subsection{What's probability: Bayesian vs Frequentist}

Probability can be interpreted as a {\bf frequency}
\begin{equation}
{\cal P}=\frac{n}{N}
\end{equation}
where $n$ stands for the successes and $N$ for the total number of trials.

Or it can be interpreted as a lack of information:  if I knew everything, I know that an event  is surely going to happen, then ${\cal P}=1$, if I know it is not going to happen then ${\cal P}=0$ but in other cases  I can use my judgment and or information from frequencies to estimate ${\cal P}$.
The world is divided in Frequentists and Bayesians. In general, Cosmologists are Bayesians and High Energy Physicists are Frequentists.

For Frequentists events are just frequencies of occurrence: probabilities are only defined as the quantities obtained in the limit when the number of independent trials tends to infinity.

Bayesians  interpret probabilities as the degree of belief in a hypothesis: they use judgment, prior information, probability theory etc...

As we do cosmology we will be Bayesian.

\subsection{Dealing with probabilities}
In probability theory, probability distributions are fundamental concepts. They are used to calculate confidence intervals, for modeling purposes etc.
We first need to introduce the concept of random variable in statistics (and in Cosmology). Depending on the problem at hand, the random variable may be the face of a dice, the number of galaxies in a volume $\delta V$ of the Universe, the CMB temperature in a given pixel of a CMB map,  the measured value of  the power spectrum $P(k)$ etc. 
The probability that $x$ (your random variable) can take a specific value is ${\cal P}(x)$ where ${\cal P}$ denotes the probability distribution.

The properties of ${\cal P}$ are:
\begin{enumerate}
\item ${\cal P}(x)$ is a  non negative, real number for all real values of $x$. 

\item ${\cal P}(x)$ is normalized so that  \footnote{ for discrete distribution $\int \longrightarrow \sum$} $\int dx{\cal  P}(x)=1$

\item For mutually exclusive events $x_1$ and $x_2$, ${\cal P}(x_1+x_2)={\cal P}(x_1)+{\cal P}(x_2)$ the probability of  $x_1$ or $x_2$  to happen is the sum of the individual probabilities. ${\cal P}(x_1+x_2) $ is also  written as ${\cal P}(x_1U x_2)$ or ${\cal P}(x_1 .OR. x_2)$.

\item In general: 
\begin{equation}
{\cal P}(a,b)={\cal P}(a){\cal P}(b|a)\,\,\,\,; \,\,\,\,\,\,\,\,\,\, {\cal P}(b,a)={\cal P}(b){\cal P}(a|b)
\label{eq:conditionalprob}
\end{equation}
 The probability of $a$ and $b$ to happen  is the probability of $a$ times the conditional probability of $b$ given  $a$. Here we can also make the (apparently tautological) identification ${\cal P}(a,b)={\cal P}(b,a)$.
For independent events then ${\cal P}(a,b)={\cal P}(a) {\cal P}(b)$.
\end{enumerate}

--------------------------------------------------------

{\bf Exercise:}  ''Will it be sunny tomorrow?"  answer in the frequentist way and in the Bayesian way\footnote{These lectures were given in the Canary Islands, in other locations answer may differ...} .\\

{\bf Exercise:}  Produce some examples for rule (iv) above.

----------------------------------------------------------

While Frequentists only consider distributions of events, Bayesians consider hypotheses  as ''events", giving us Bayes theorem:
 \begin{equation}
 {\cal P}(H|D)=\frac{{\cal P}(H){\cal P}(D|H)}{{\cal P}(D)}
 \label{eq:bayes}
 \end{equation}
 where $H$  stands for hypothesis (generally the set of parameters specifying your model, although many cosmologists now also consider model themselves) and $D$ stands for data.  ${\cal P}(H|D)$ is called the {\bf posterior} distribution. ${\cal P}(H)$ is called the {\bf prior} and ${\cal P}(D|H)$ is called {\bf likelihood}. 
 
 Note that  this is nothing but equation \ref{eq:conditionalprob} with the apparently tautological identity ${\cal P}(a,b)={\cal P}(b,a)$ and with substitutions: $b \longrightarrow H$ and $a\longrightarrow D$.
 
 Despite its simplicity  Eq. \ref{eq:bayes} is a really important equation!!!

 The usual points of heated discussion follow: ''How do you chose ${\cal P}(H)$?",  ''Does the choice affects your final results?" (yes, in general it will).  ''Isn't this then a bit subjective?"

--------------------------------------------------------

{\bf Exercise:}   Consider a  positive definite quantity (like for example the tensor to scalar ratio $r$ or the optical depth to the last scattering surface $\tau$). What prior should one use? a flat prior in the variable? or a logaritmic prior (i.e. flat prior in the log of the quantity)?  for example CMB analysis may use a flat prior in $\ln r$, and in $Z=\exp(-2 \tau)$. How is this related to using a flat pror in $r$ or in $\tau$?
 It will be useful to consider the following:
 effectively we are comparing ${\cal P}(x)$ with ${\cal P}(f(x))$, where $f$ denotes a  function of $x$.  For example $x$ is $\tau$ and $f(x)$ is $\exp(-2 \tau)$.
 Recall that:
 ${\cal P}(f)={\cal P}(x(f))\left| \frac{df}{dx}\right|^{-1}$. The Jacobian of the transformation appears here to conserve probabilities.
 
{\bf Exercise:} Compare Fig 21  of Spergel et al (2007)\cite{spergel07} with figure 13  of Spergel et al 2003 \cite{spergel03}. Consider the WMAP only contours. Clearly the 2007 paper  uses more data than the 2003 paper, so why is that the constraints look worst? If you suspect the prior you are correct! Find out which prior has changed, and why it makes such a difference.

{\bf Exercise:} Under which conditions the choice of prior does not matter? (hint: compare the  WMAP papers of 2003 and 2007 for the  flat LCDM case).

--------------------------------------------------------\\

\subsection{Moments and cumulants}
Moments and cumulants are used to characterize the probability distribution.
 In the language of probability distribution {\bf averages} are defined as follows:
 \begin{equation}
 \langle f(x) \rangle=\int dx f(x) {\cal P}(x)
 \label{eq:average}
 \end{equation}
 These can then be related to "expectation values" (see later).
 For now let us just  introduce the moments: $\hat{\mu}_m =\langle x^m\rangle$ and, of special interest, the central moments: $\mu_m=\langle (x-\langle x\rangle)^m\rangle$.
 
 ----------------------------------------------------------------
 
{\bf Exercise}: show that $\hat{\mu}_0=1$ and that the average $\langle x\rangle=\hat{\mu}_1$. Also show that $\mu_2=\langle x^2\rangle-\langle x\rangle^2$
 
 ------------------------------------------------------------------
 
 Here, $\mu_2$ is the variance, $\mu_3$ is called the skewness, $\mu_4$ is related to the kurtosis. If you deal with the statistical nature of initial conditions (i.e. primordial non-Gaussianity) or non-linear evolution of Gaussian initial conditions, you will encounter these quantities again (and again..).

Up to the skewness,  central moments and cumulants coincide. For higher-order terms things become more complicated.  To keep things a simple as possible  let's just consider  the Gaussian distribution  (see below) as reference.  While moments of order higher than 3  are  non-zero  for both Gaussian and non-Gaussian distribution,  the {\bf cumulants}  of higher orders are zero for a Gaussian distribution. In fact, for a Gaussian distribution all moments of order higher than 2 are specified by $\mu_1$ and $\mu_2$. Or, in other words, the mean and the variance  completely specify a Gaussian  distribution. This is not the case for a non-Gaussan  distribution.  For non-Gaussian distribution,
the relation between central moments    and cumulants $\kappa$ for the first 6 orders is reported below.
\begin{eqnarray}
\mu_1&=&0\\
\mu_2&=&\kappa_2\\
\mu_3&=&\kappa_3 \\
\mu_4&=&\kappa_4+3 (\kappa_2)^2\\
\mu_5&=&\kappa_5+10\kappa_3\kappa_2\\
\mu_6&=&\kappa_6+15\kappa_4\kappa_2+10(\kappa_3)^2+15(\kappa_2)^3
\end{eqnarray}

 \subsection{Useful trick: the generating function}
 The generating function allows one, among other things,  to  compute quickly moments and cumulants of a distribution.
 Define the generating function as
 \begin{equation}
 Z(k)=\langle \exp(i k x)\rangle = \int dx \exp(i k x) {\cal P}(x)
 \label{eq:genfunc1}
 \end{equation}
 Which may sound familiar as it is a sort of Fourier transform...
 Note that this can be written as an infinite series (by expanding the exponential) giving (exercise)
 \begin{equation}
 Z(k)=\sum_{n=0}^{\infty}\frac{(ik)^n}{n!} \hat{\mu}_n
  \end{equation}
 
 So far nothing special, but now the neat trick is that the {\bf moments} are obtained as:
\begin{equation}
 {\hat \mu}_n=(-i^n) \frac{d^n}{dk^n} Z(k)|_{k=0}
 \end{equation}
and the {\bf cumulants} are obtained  by doing the same operation on $\ln Z$.
While this seems just a neat trick for now it will be very useful shortly.

 \subsection{Two  useful distributions}
 Two distributions are widely used in Cosmology: these are the Poisson distribution and the Gaussian distribution.
 
 \subsubsection{The Poisson distribution}
 The Poisson distribution describes an independent point process:  photon noise, radioactive decay, galaxy distribution for very few galaxies, point sources ....
 It is an example of a discrete probability distribution.
 For cosmological applications it is useful to think of a Poisson process as follows.  Consider a random process (for example a random distribution of galaxies in space) of average density $\rho$. 
 Divide the space in infinitesimal cells, of volume $\delta V$ so small that their occupation can only be $0$ or $1$ and the probability of having more than one object per cell is $0$. Then the probability of having one object  in a given cell is ${\cal P}_1=\rho \delta V$ and the probability of getting no object in the cell is therefore ${\cal P}_0=1-\rho \delta V$. 
 Thus for one cell  the generating function is $Z(k)=\sum_n{\cal P}_n\exp(i k n)=1+\rho \delta V (\exp(ik)-1)$ and for  a volume $V$ with $V/\delta V$ cells,  we have $Z(k)=(1+\rho \delta V (\exp(ik)-1))^{V/\delta V} \sim \exp[\rho V (\exp(ik)-1)]$.
 
 With the substitution $\rho V \longrightarrow \lambda$ we obtain \\
 $Z(k)=\exp[\lambda(\exp(ik)-1)]=\sum_{n=0}^{\infty}\lambda^n/n! \exp(-\lambda)\exp(ikn)$.
 Thus the Poission probability distribution  we recover is:
 \begin{equation}
 {\cal P}_n=\frac{\lambda^n}{n!}\exp[-\lambda]
 \end{equation}

 -----------------------------------------------------------------------------------------\\
 Exercise:  show that for the Poisson distribution $\langle n \rangle=\lambda$ and that $\sigma^2=\lambda$.
 
 ----------------------------------------------------------------------------------------\\

\subsubsection{The Gaussian distribution} 
 The Gaussian distribution is extremely useful because of the ''Central Limit theorem". The Central Limit  theorem states that  the sum of many independent and identically distributed random variables  will be approximately Gaussianly distributed. The conditions for this to happen are quite mild: the variance of the distribution one starts off with has to be finite.
The proof is remarkably simple.
 Let's take  $n$ events with probability distributions ${\cal P}(x_i)$ and $<x_i>=0 $ for simplicity,  and let 
Y  be  their sum. What is ${\cal P}(Y)$?  The generating function for $Y$ is the product of the 
generating functions for the $x_i$:
  \begin{equation}
  Z_Y(k)=\sum_{m=0}^{m=\infty}\left[\frac{(ik)_m}{m!}\mu^m\right]^n\simeq \left(1-\frac{1}{2} \frac{k^2<x^2>}{n}+...\right)^n
  \end{equation}  
  for $n\longrightarrow \infty$ then $Z_Y(k)\longrightarrow \exp[-1/2k^2 <x^2>]$.
  By recalling the definition of generating function  (eq. \ref{eq:genfunc1}
 we can see that the probability distribution which generated this $Z$ is 
 \begin{equation}
  {\cal P}(Y)=\frac{1}{\sqrt{2 \pi <x^2>}}\exp\left[-\frac{1}{2} \frac{Y^2}{<x^2>}\right]
 \end{equation}
 that is  a Gaussian!

 ----------------------------------------------------------------------------------------------- 
 
{\bf  Exercise}: Verify that  higher order cumulants are zero for the Gaussian distribution.\\

 {\bf Exercise}: Show that the Central limit theorem holds for the  Poisson distribution.
 
------------------------------------------------------------------------------------------------

 Beyond the Central Limit theorem, the Gaussian distribution   is very important in cosmology as we believe that the initial conditions, the primordial perturbations generated from inflation, had a distribution very very close to Gaussian. (Although it is crucial to test this experimentally.)

We should also remember that thanks to the Central Limit theorem, when we  estimate parameters in cosmology in many cases we approximate  our data as having a Gaussian distribution, even if we know that each data point is NOT drawn from a Gaussian distribution. The Central Limit theorem simplifies our lives every day...

There are exceptions though. Let us  for example consider $N$ independent data points drawn from a Cauchy distribution: ${\cal P}(x)=[\pi \sigma (1+[(x-\bar{x})/\sigma]^2]^{-1}$. This is a proper probability distribution as it integrates to unity, but moments diverge. One can show that the numerical mean of a finite number $N$ of observations is finite but the "population mean" (the one defined through  the integral of equation (\ref{eq:average}) with $f(x)=x$) is not.  Note also that the scatter in the average of $N$ data points drawn from this distribution is the same as the scatter in 1 point: the scatter never diminishes regardless of the sample size....

\section{Modeling of data and statistical inference}
To illustrate this let us follow the example from \cite{statastro}.
If you have an urn with $N$ red balls and $M$ blue balls and you draw from the urn, probability theory can tell you what  the chances are of you to pick a red ball given that you has so far drawn $m$ blue and $n$ red ones... However in practice what you want to do is to use probability to tell you what is the distribution of the balls in the urn having made a few drawn from it!

In other words, if you knew everything about the Universe, probability theory could tell you what the probabilities are to get a given outcome  for an observation. However, especially in cosmology, you want to make few observations and draw conclusions about the Universe! With the added complication that   experiments in Cosmology  are not quite like experiments in the lab: you can't poke the Universe and see how it reacts, and in many cases you can't repeat the observation, and you can only see a small part of the Universe! Keeping this caveat in mind let's push ahead.

Given a set of observations often you want to fit a model to the data, where the model is described by a set of parameters $\vec{\alpha}$. Sometimes the model is physically motivated (say CMB angular power spectra etc.) or a convenient function (e.g. initial studies of large scale structure were fitting galaxies correlation functions with power laws). Then you want to define a merit function,  that measures the agreement between the data and the model: by adjusting the parameters to  maximize the agreement  one obtains the {\it best fit parameters}. Of course, because of measurement errors, there will be errors associated to the parameter determination.
 To be useful  a fitting procedure  should provide {\it a)} best fit parameters {\it b)} error estimates on the parameters {\it c)} possibly a statistical measure of the goodness of fit.  When {\it c)} suggests that the model is a bad  description of the data, then {\it a)} and {\it b)} make no sense.
 
Remember at this point Bayes theorem: while  you may want to ask: ''What is the  probability that a particular set of parameters  is correct?", what  you can ask to a {\it ''figure of merit"} is  ''Given a set of parameters, what is the probability that  that this data set could have occurred?". This is the likelihood. 
 You may want to estimate parameters by maximizing the likelihood and somehow identify the likelihood (probability of the data given the parameters) with the likelihood of the model parameters.    
    
\subsection{Chisquare, goodness of fit and confidence regions}
Following Numerical recipes (\cite{numrec}, Chapter 15)  it is easier to introduce model fitting and parameter estimation using the least-squares example. 
Let's say that $D_i$ are our data points and $y(\vec{x_i}|\vec{\alpha})$  a model  with parameters $\vec{\alpha}$ . For example if the model is a straight line then $\vec{\alpha}$ denotes the slope and intercept of the line.

\begin{figure}
\begin{center}
\setlength{\unitlength}{1mm}
\begin{picture}(000,130)
\includegraphics{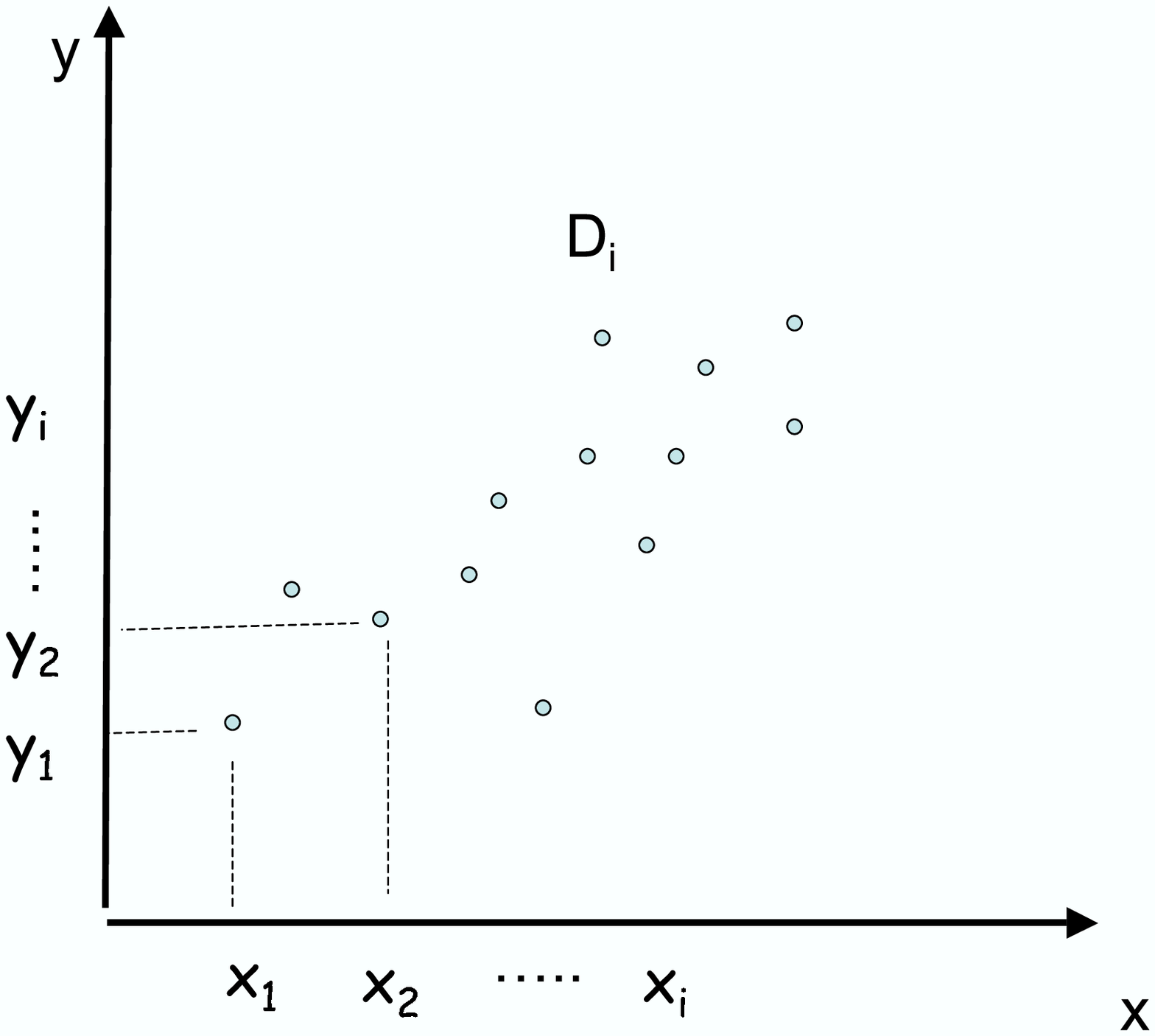}
\end{picture}
\end{center}
\caption{Example of linear fit to the data points $D_i$ in 2D.}
\label{fig:1}
\end{figure}

 The least squares is given by:
 \begin{equation}
 \chi^2=\sum_i w_i[D_i-y(x_i|{\vec{\alpha}})]^2
 \end{equation}
 and you can show that the minimum variance weights are $w_i=1/\sigma^2_1$. 
 
 -------------------------------------------------------------------------------
 
{\bf  Exercise}: if the  points are correlated how does this equation change?

 ----------------------------------------------------------------------------------

  Best fit value parameters are the parameters that minimize the $\chi^2$. 
Note that a numerical exploration can be avoided:  by solving $\partial \chi^2/\partial \alpha_i \equiv 0$ you can find the best fit parameters.
 
  \subsubsection{Goodness of fit}
  In particular, if the measurement errors are Gaussianly distributed,  and (as in this example)  the model is a linear function of the parameters, then the probability distribution of for different values of $\chi^2$ at the minimum is the $\chi^2$ distribution for $\nu \equiv n-m$ degrees of freedom (where $m$ is the number of parameters and $n$ is the number of data points.  The probability that the observed $\chi^2$ even for a correct model is less than a value $\hat{\chi}^2$ is ${\cal P}(\chi^2<\hat{\chi}^2, \nu)={\cal P}(\nu/2,\hat{\chi}^2/2)=\Gamma(\nu/2,\hat{\chi}^2/2)$ where $\Gamma$ stands for the incomplete  Gamma function. Its complement, $Q=1-{\cal P}(\nu/2,\hat{\chi}^2/2)$ is the probability that the observed $\chi^2$ exceed  by chance  $\hat{\chi^2}$ even for a correct model. See numerical recipes \cite{numrec} chapters 6.3 and 15.2 for more details.
  It is common that the chi-square distribution holds even for models that are non linear in the parameters and even in more general cases (see an example later).
  
  The computed probability $Q$ gives a quantitative measure of the goodness of fit when evaluated at the best fit parameters (i.e. at $\chi^2_{min}$). If $Q$ is a very small probability  then 
\begin{itemize}
\item[]{\it a)} the model is wrong and can be rejected 
\item[]{\it b)} the errors are really larger than stated or 
\item[]{\it c)} the measurement errors were not Gaussianly distributed. 
\end{itemize}
If you know the actual error distribution you may want to {\bf Monte Carlo simulate}  synthetic data stets, subject them to your actual fitting procedure,  and determine both the probability distribution of your $\chi^2$ statistic and the accuracy with which model parameters are recovered by the fit (see section on Monte Carlo Methods).

On the other hand $Q$ may be too large, if it is too near 1  then also something's up: 
\begin{itemize}
\item[]{\it a)}  errors may have been overestimated
\item[]{\it b)}  the data are correlated and correlations were ignored in the fit. 
\item[]{\it c)} In principle it may be that the distribution you are dealing with is more compact than a Gaussian distribution, but this is almost never the  case. So make sure you exclude cases {\it a)} and {\it b)} before you invest a lot of time in exploring option {\it c)}.
\end{itemize}

Postscript: the ''Chi-by eye" rule is that the minimum $\chi^2$ should be roughly equal to the number of data-number of parameters (giving rise to the widespread use of the so-called reduced chisquare). Can you --possibly rigorously-- justify this statement?

\subsubsection{Confidence region}
Rather than presenting the full probability distribution of errors it is useful to present confidence limits or confidence regions: a region in the $m$-dimensional parameter  space ($m$ being the number of parameters), that contain a certain percentage of the total probability distribution. 
Obviously you want a suitably compact region around the best fit value. It is customary to choose 68.3\%, 95.4\%, 99.7\%... Ellipsoidal regions have connections with the normal (Gaussian) distribution but in general  things may be very different...
A natural choice for the shape of confidence intervals is given by constant $\chi^2$ boundaries.
For the observed data set the value of parameters $\vec{\alpha}_0$ minimize the $\chi^2$, denoted by  $\chi^2_{min}$. If we perturb $\vec{\alpha}$ away from $\vec{\alpha}_0$ the $\chi^2$ will increase.
From the properties of the $\chi^2$ distribution it is possible to show that 
there is a well defined relation between  confidence intervals, formal standard errors, and $\Delta\chi^2$. We In table\ref{table:chisq}  we report  the $\Delta \chi^2$ for  the conventionals $1,2$,and $3-\sigma$ as a function of the number of parameters for the joint confidence levels.

In general, let's spell out the following prescription.
If $\mu$ is the number of fitted parameters for which you want to plot the join confidence region and $p$ is the confidence limit desired,  find the $\Delta \chi^2$ such that the probability of a chi-square variable with $\mu$ degrees of freedom being less than $\Delta \chi^2$ is $p$. For general  values of $p$ this is given by $Q$ described above (for the standard 1,2,3$-\sigma$ see table above).

P.S. Frequentists use $\chi^2$ a lot.

 \subsection{Likelihoods}
One can  be more sophysticated than $\chi^2$, if ${\cal P}(D)$ (D is data) is known.
 Remember from the Bayes theorem (eq.\ref{eq:bayes})  the probability of the data given the model (Hypothesis)  is  the likelihood. If we set ${\cal P}(D)=1$ (after all, you  got the data) and ignore the prior, 
by maximizing the likelihood we find the most likely Hypothesis, or , often,  the most likely parameters of a given model.
 
 Note that we have ignored ${\cal P}(D)$ and the prior so in general this technique does not give you a goodness of fit  and not an absolute probability of the model, only relative probabilities. Frequentists rely on $\chi^2$ analyses where a goodness of fit can be established. 

 In many cases (thanks to the central limit theorem) the likelihood can be well approximated by a multi-variate Gaussian:
 
 \begin{equation}
 {\cal L}=\frac{1}{(2\pi)^{n/2} |det C|^{1/2}} \exp\left[-\frac{1}{2} \sum _{ij}(D-y)_iC^{-1}_{ij}(D-y)_j\right]
 \end{equation}
 
 where $C_{ij}=\langle (D_i-y_i)(D_j-y_j)\rangle$ is the covariance matrix.

 ---------------------------------------------------------------------------------------
 
 Exercise:  when are likelihood analyses and $\chi^2$ analyses the same?
 
 -----------------------------------------------------------------------------------------

 \subsubsection{Confidence levels for likelihood}
  For Bayesian statistics,  confidence regions are found as regions $R$ in {\it model space}  such that $\int_R {\cal P}({\vec{\alpha}}|D)d\vec{\alpha}$ is, say, 0.68 for 68\% confidence level and  0.95 for 95\% confidence.   Note that this encloses the prior information. To report results independently of the prior the likelihood ratio is used. In this case compare the likelihood at a particular point in model space ${\cal L}(\vec{\alpha})$ with the value of the maximum likelihood ${\cal L}_{max}$. Then a model is said acceptable if 
  \begin{equation}
-2\ln \left[\frac{{\cal L}(\vec{\alpha})}{{\cal L}_{max}}\right] \le {\rm threshold}
  \end{equation}
  Then the threshold should be calibrated by calculating the distribution of the likelihood ratio  in the case where a particular model  is the true model. 
  There are some cases however when the value of the threshold is the corresponding confidence limit for a $\chi^2$ with $m$ degrees of freedom, for $m$ the number of parameters.  
  
  -----------------------------------------------------------------------------------------
  
  {\bf Exercise}: in what cases?\footnote{Solution: The data must have Gaussian errors, the model must depend  linearly on the parameters, the gradients of the model with respect to  the parameters are not degenerate and  the parameters do not affect the covariance.}
  
  ---------------------------------------------------------------------------------------------

 \subsection{Marginalization, combining different experiments}
 Of all the model parameters $\alpha_i$ some of them may be uninteresting. Typical  examples  of nuisance parameters  are calibration factors, galaxy bias parameter etc,  but also it may be that we are interested on constraints on  only one cosmological parameter at the time rather than on the {\it joint}  constraints on 2 or more parameters simultaneously. 
 One then marginalizes over the uninteresting parameters by integrating the posterior distribution:
 \begin{equation}
 P(\alpha_1..\alpha_j|D)=\int d\alpha_{j+1},...d \alpha_{m} P(\vec{\alpha}|D)
 \end{equation}
 if there are in total $m$ parameters and we are interested in $j$ of them ($j<m$).
Note that if you have two independent experiments, the combined likelihood of the two experiments is just the product of the two likelihoods. (of course if the two experiments are non independent then one would have to include their covariance). In many cases one of the two experiments can be used as a prior. 
 A word of caution is  on order here.  We can always combine independent experiments by multiplying their likelihoods,   and if the experiments are good  and sound and the model used is a good and complete description of the data  all is well. However it is always important to: {\it a)} think about the priors one is using and to  quantify their effects. {\it  b)}  make sure that results from independent experiments are consistent: by multiplying likelihood from inconsistent experiments you can always get some sort of results  but it does not mean that the result actually makes sense....
 
   Sometimes you may be interested in placing a prior on the uninteresting parameters before marginalization. The prior may  come from a previous measurement or from your "belief".

Typical examples of this are: marginalization over calibration uncertainty, over point sources amplitude  or over beam errors  for CMB studies. For example for marginalization over, say, point source amplitude, it is useful to know of the following trick for Gaussian  likelihoods:
\begin{eqnarray}
P(\alpha_1..\alpha_{m-1}|D)\!\!\!\!&=&\!\!\!\!\!\!\int \frac{dA}{(2\pi)^{\frac{m}{2}} ||C||^{\frac{1}{2}}} e^{\left[-\frac{1}{2}(C_i-(\hat{C_i}+AP_i))\Sigma_{ij}^{-1}(C_j-(\hat{C_j}+AP_j))\right]}  \\ \nonumber
&\times&\frac{1}{\sqrt{2\pi \sigma^2}}\exp\left[-\frac{1}{2}\frac{(A-\hat{A})^2}{\sigma^2}\right]         
\end{eqnarray}
repeated indices are summed over and $||C||$ denotes the determinant. Here, $A$ is the amplitude of, say, a point source contribution $P$ to the $C_{\ell}$ angular power spectrum, $A$ is the  $m-th$ parameter which we want to marginalize over  with a  Gaussian prior with variance $\sigma^2$ around $\hat{A}$.
The trick is to recognize that this    integral can be written as:
\begin{equation}
\!\!\!\!P(\alpha_1..\alpha_{m-1}|D)=C_0 \exp\left[-\frac{1}{2}C_1- 2 C_2 A+C_3 A^2\right]dA
\end{equation}
(where $C_{0...3}$ denote constants)
and that  this kind of integral  is evaluated by using the substitution $A\longrightarrow A-C_2/C_3$ giving
something $\propto \exp[-1/2(C_1 - C_2^2/C_3)]$.

It is left as an exercise to write the constants  explicitly.

\subsection{An example}
Let's say you want to constrain cosmology by studying clusters number counts as a function of redshift. Here we follow the paper of Cash 1979 \cite{cash79}.
The observation of a discrete number $N$ of clusters is  a Poisson process, the probability of which is given by the product 
\begin{equation}
{\cal P}=\Pi_{i=1}^{N}[e_i^{n_i}\exp(-e_i)/n_i!]
\end{equation}
 where $n_i$ is the number of clusters observed in the $i-th$
 experimental bin and $e_i$ is the expected number in that bin in a
 given model: $e_i=I(x)\delta x_i$ with $i$ being the proportional to
 the probability distribution.  Here $\delta x_i$ can represent an interval in clusters mass and/or redshift.
 Note: this is a product of Poisson distributions, thus one is assuming that these are independent processes. Clusters may be clustered, so when can this be used?
 
 For unbinned data (or for small bins so
 that bins have only 0 and 1 counts) we define the quantity:
\begin{equation}
C\equiv -2 \ln {\cal P}=2(E-\sum_{i=1}^N\ln I_i)
\end{equation}
where $E$ is the total expected number of clusters in a given  model.
The quantity $\Delta C$ between two models with different parameters has a $\chi^2$ distribution! (so all  that was said in the $\chi^2$ section applies, even though we started from a highly non-Gaussian distribution.)

 \section{Description of random fields}
 
 Let's take a break from probabilities and consider a slightly different issue.
 In comparing the results of theoretical calculations with the observed
Universe, it would be meaningless to hope to be able to describe with the
theory the properties of a particular patch, i.e. to predict the density
contrast of the matter $\delta (\vec{x})=\delta \rho(x)/\rho $ at any specific
point $\vec{x}$. Instead,
it is possible to predict the average statistical properties of the mass
distribution \footnote{A very similar approach is taken in statistical
mechanics.}. 
In addition, we consider that  the Universe we live in is a random realization of all the possible Universes that could have been a realization of the  true underlying model (which is known only to Mother Nature).  All the possible realizations of this true underlying Universe   make up the {\it ensamble}. In statistical inference one may sometime  want to try to estimate how different our particular realization of the Universe  could be from the true underlying one.  Thinking back at the  example of the urn with colored balls, it would be like considering that the particular urn from which we are drawing the balls is only one possible realization of the true underlying distribution of urns. For example, say that the true distribution has a 50-50 split in  red and blue balls but that the urn can have only an odd number of balls. Clearly the exact 50-50 split cannot be realized in one particular urn but it can be realized in the ensamble...

 Following the {\it cosmological principle} (e.g. Peebles 1980 \cite{Peebles80}),
models of the Universe have to be homogeneous on the average, therefore, in
widely separated regions of the Universe (i.e.  independent), the density
field must have the same statistical properties.

A crucial assumption of standard cosmology is that the part of the Universe
that we can observe is a {\it fair sample} of the whole.
This is closely
related to the {\it cosmological principle} since it implies that the
statistics like the correlation functions have to be considered as averages
over the ensemble.  But the peculiarity in cosmology is that we have just one
Universe, which is just one realization from the ensemble (quite fictitious
one: it is the ensemble of all possible Universes). The fair sample
hypothesis states that samples from well separated part of the Universe are
independent realizations of the same physical process, and that, in the
observable part of the Universe, there are enough independent samples to be
representative of the statistical ensemble.  The hypothesis of ergodicity
follows: averaging over many realizations is equivalent to averaging over a
large (enough) volume.
The cosmological field we are interested in, in a given volume, is taken as a realization of the statistical 
process and, for the hypothesis of ergodicity, averaging over many 
realizations is equivalent to averaging over a large volume.

Theories can just predict the statistical properties of $\delta (\vec{x})$
which, for the cosmological principle, must be a homogeneous and isotropic random field, and our
observable Universe is a random realization from the ensemble.

In cosmology the scalar field $\delta (\vec{x})$ is enough to specify the initial
fluctuations field, and --we  ultimately hope-- also the present day distribution of
galaxies and matter. Here lies one of the big  challenges of modern cosmology. (see e.g. \ref{MS02})
 
A fundamental problem in the analysis of the cosmic structures, is to find the
appropriate tools to provide information on the distribution of the density
fluctuations, on their initial conditions and subsequent evolution.
Here we concentrate on power spectra and correlation functions.

\subsection{Gaussian random fields}
 Gaussian random fields are crucially important in cosmology, for different
reasons: first of all it is possible to describe their statistical properties
analytically, but also there are strong theoretical motivations, namely inflation, to assume that
the primordial fluctuations that gave rise to the present-day cosmological
structures, follow a Gaussian distribution. Without resorting to inflation, for
the central limit theorem, Gaussianity results from  a superposition of a large
number of random processes.

The distribution of density fluctuations $\delta$ defined as \footnote{Note that $\langle \delta \rangle=0$}
$\delta=\delta\rho /\rho$ cannot be exactly Gaussian because the field has to
satisfy the constraint $\delta>-1$, however if the amplitude of the
fluctuations is small enough, this can be a good approximation. This seems
indeed to be the case: by looking at the CMB anisotropies we can probe
fluctuations when their statistical distribution should have been close to its
primordial one; possible deviations from Gaussianity of the primordial density field  are small.

If $\delta$ is a Gaussian random field with average 0, its probability distribution is given
by:
\begin{equation}
P_n(\delta_1,\cdot\cdot\cdot,\delta_n)=\frac{\sqrt{\mbox{Det}{\bf
C}^{-1}}}{(2\pi)^{n/2}}\exp\left[-\frac{1}{2} {\bf \delta}^T {\bf
C}^{-1} {\bf \delta} \right]
\end{equation}
where ${\bf \delta}$ is a vector made by the $\delta_i$, ${\bf C}^{-1}$ denotes
the inverse of the 
correlation matrix which elements are ${\bf C}_{ij}=\langle\delta_i\delta_j \rangle$.

An important property of Gaussian random fields is that the Fourier transform
of a Gaussian  field is still Gaussian. The phases of the Fourier modes are
random and the real and imaginary
part of the coefficients have Gaussian distribution and are mutually
independent. 

Let us denote the real and imaginary part of $\delta_{\bf k}$ by $Re\delta_{\bf
k}$ and $Im\delta_{\bf k}$ respectively. 
Their joint probability distribution is the bivariate Gaussian:
\begin{equation}
P(Re\delta_{\bf k}, Im\delta_{\bf k})dRe\delta_{\bf k} dIm\delta_{\bf k}
=\frac{1}{2\pi \sigma^2_k}\exp\left[-\frac{Re\delta_{\bf k}^2+Im\delta_{\bf
k}^2}{2 \sigma^2_k}\right]  dRe\delta_{\bf k} dIm\delta_{\bf k}
\label{eq:reimgau}
\end{equation}
where $\sigma^2_k$ is the variance in $Re\delta_{\bf k}$ and $Im\delta_{\bf
k}$ and  for isotropy it depends only on the magnitude of ${\bf k}$.
Equation (\ref{eq:reimgau}) can be re-written in terms of the amplitude
$|\delta_{\bf k}|$ and the phase $\phi_{\bf k}$:
\begin{equation}
P(|\delta_{\bf k}|,\phi_{\bf k})d |\delta_{\bf k}| d \phi_{\bf
k}=\frac{1}{2\pi\sigma_k^2}\exp\left[-\frac{|\delta_{\bf k}|^2}{2\sigma_k^2}
\right]|\delta_{\bf k}| d |\delta_{\bf k}| d \phi_{\bf k}
\end{equation}
that is  $|\delta_k|$ follows a Rayleigh distribution.

From this follows that the probability that the amplitude is above a certain
threshold $X$ is:
\begin{equation}
P(|\delta_{\bf k}|^2 >
X)=\int_{\sqrt{X}}^{\infty}\frac{1}{\sigma_k^2}\exp\left[-\frac{|\delta_{\bf
k}|^2}{2\sigma_k^2}\right]|\delta_{\bf k}| d |\delta_{\bf
k}|=\exp\left[-\frac{X}{\langle |\delta_{\bf k}|^2 \rangle}\right]\;.
\end{equation}
Which is  an exponential distribution.

The fact that the phases of a Gaussian field are random, implies that the two
point correlation function (or the power spectrum) completely specifies the field. 

-----------------------------------------------------------------------------------------\\

P.S. If your advisor now asks you to generate a Gaussian random field you know how to do it.
(It you are not familiar with Fourier transforms see next section)

-------------------------------------------------------------------------------------------\\

The observed fluctuation field however is not Gaussian.  The observed galaxy
distribution is highly non-Gaussian  principally due to gravitational
instability.  To completely specify a
non-Gaussian distribution higher order correlation functions are
needed\footnote{For ``non pathological'' distributions. For a discussion see
e.g. \cite{kendallstewardt77}.};
conversely  deviations from Gaussian behavior can be characterized by the higher-order statistics of the distribution.

 \subsection{Basic tools}

The Fourier transform of the (fractional) overdensity field $\delta$ is defined as:
\begin{equation}
\delta_{\vec{k}}=A\int d^3r \delta(\vec{r}) \exp[- i \vec{k}\cdot \vec{r}]
\end{equation}
with inverse
\begin{equation}
\delta(\vec{r})=B\int d^3k \delta_{\vec {k}}\exp[i \vec{k}\cdot\vec{r}]
\end{equation}
 and the Dirac delta is then given by 
 \begin{equation}
 \delta^D(\vec{k})=BA\int d^3r \exp[\pm i\vec{k}\cdot \vec{r}]
 \end{equation}
Here I chose the convention $A=1$, $B=1/(2 \pi)^3$, but always beware of the FT conventions. 

The  two point {\bf correlation function} (or correlation function) is defined as:
\begin{equation}
\xi(x)=\langle \delta(\vec{r})\delta(\vec{r}+\vec{x})\rangle=\int <\delta_{\vec{k}} \delta_{\vec{k'}}> \exp[i\vec{k}\cdot \vec{r}]\exp[i \vec{k} \cdot (\vec{r}+\vec{x})] d^3k d^3k'
\end{equation}
because of isotropy $\xi(|x|)$ (only a function of the distance not orientation). Note that in some cases when isotropy is broken one may want to keep the orientation information (see e.g. redshift space distortions, which affect clustering only along the line-of sight ).

The definition of the power spectrum $P(k)$ follows :
\begin{equation}
<\delta_{\vec{k}}\delta_{\vec{k'}}>=(2\pi)^3 P(k) \delta^D(\vec{k}+\vec{k'})
\end{equation}
again  for isotropy $P(k)$ depends only on the modulus of the k-vector, although in special cases where isotropy is broken one may want to keep the direction information.

Since $\delta(\vec{r})$ is real. we have that $\delta_{\vec{k}}^*=\delta_{-\vec{k}}$, so
\begin{equation}
<\delta_{\vec{k}} \delta^*_{\vec{k'}} >= (2\pi)^3 \int d^3x \xi(x) \exp[-i\vec{k}\cdot \vec{x}] \delta^d(\vec{k}-\vec{k'})
\end{equation}\'.

The power spectrum and the correlation function are Fourier transform pairs:
\begin{eqnarray}
\xi(x)&=&\frac{1}{(2\pi)^3}\int P(k)\exp[i\vec{k}\cdot \vec{r}]d^3k\\
P(k)&=&\int \xi(x)\exp[-i\vec{k}\cdot \vec{x}]d^3x
\end{eqnarray}
At this stage the same amount of information is enclosed in $P(k)$ as in $\xi(x)$.

From here the variance is
\begin{equation}
\sigma^2=<\delta^2(x)>=\xi(0)=\frac{1}{(2\pi)^3}\int P(k) d^3k
\end{equation} 
or better
\begin{equation}
\sigma^2=\int \Delta^2(k)d\ln k \,\,{\rm where}\,\,\Delta^2(k)=\frac{1}{(2 \pi)^3}k^3 P(k)
\end{equation}
and the quantity $\Delta^2(k)$ is independent form the FT convention used.

Now the question is: on what scale is  this variance defined?

Answer: in practice one needs to use filters: the density field is convolved with a filter (smoothing) function. There are two typical choices:
\begin{equation}
f=\frac{1}{(2\pi)^{3/2}R_G^3}\exp[-1/2 x^2/R_G^2] \,\,\,\,{\rm Gaussian} \rightarrow f_k=\exp[-k^2R_G^2/2]
\end{equation}

\begin{equation}
f=\frac{1}{(4\pi)R_T^3}\Theta(x/R_T) \,\,\,\,{\rm Top Hat} \rightarrow f_k=\frac{3}{(k R_T)^3}[\sin(k R_T)-k R_T\cos(k R_T)]
\end{equation}

roughly $R_T\simeq \sqrt{5}R_G$.

Remember: {\bf Convolution in real space is a multiplication in Fourier space}; {\bf Multiplication in real space is a convolution in Fourier space}.

----------------------------------------------------------------------------------\\
Exercise: consider a multi-variate Gaussian distribution:
\begin{equation}
P(\delta_1..\delta_n)=\frac{1}{(2\pi)^{n/2}{\rm det} {\bf C}^{1/2}}\exp[-\frac{1}{2}\delta^T {\bf C}^{-1}\delta]
\end{equation}
where $C_{ij}=<\delta_i \delta_j>$ is the covariance. 
Show that is $\delta_i$ are Fourier modes then $C_{ij}$ is diagonal.
This is an ideal case, of course but this is telling us that for Gaussian fields the different $k$ modes are independent! which is always a nice feature.

Another question for you:  if you start off with a Gaussian distribution (say from Inflation) and then leave this Gaussian field $\delta$ to evolve under gravity, will it remain Gaussian forever?
Hint: think about present-time Universe, and think about the dark matter density at, say, the center of a big galaxy and in a large void.

 -----------------------------------------------------------------------------------------\\

\subsubsection{The importance of the Power spectrum}
The structure of the Universe on large scales is largely dominated by the force of gravity (which we think we know well) and no too much by complex mechanisms (baryonic physics, galaxy formation etc.)- or at least that's the hope... 
Theory (see lectures on inflation) give us a prediction for the primordial power spectrum:
\begin{equation}
P(k)=A\left(\frac{k}{k_0}\right)^n
\end{equation}
$n$ - the {\bf spectral index} is often taken to be a constant and the power spectrum is a power law power spectrum. However there are theoretical motivations to generalize this to
\begin{equation}
P(k)=A\left(\frac{k}{k_0}\right)^{n(k_0)+\frac{1}{2}\frac{dn}{d\ln k}\ln(k/k_0)}
\end{equation}
as a sort of Taylor expansion of $n(k)$ around the pivot point $k_0$. $dn/d\ln k$ is called the {\bf running of the spectral index}.

 Note that  different authors often use different choices of $k_0$ (sometimes the same author in the same paper uses different choices...) so things may get confused.... so let's report  explicitly the conversions:
\begin{equation}
A(k_1)=A(k_0)\left(\frac{k}{k_0}\right)^{n(k_0)+1/2 (dn/d\ln k)\ln (k_1/k_0)}
\end{equation}

-----------------------------------------------------------------------------------------

{\bf Exercise:} Prove the equations above.

{\bf Exercise}:Show that given the above definition of the running of the spectral index, $n(k)=n(k_0)+dn/d\ln k \ln(k/k_0)$.

-----------------------------------------------------------------------------------------------

It can be shown that as long as linear theory applies --and only gravity is at play-- $\delta <<1$,  different Fourier modes evolve independently  and the Gaussian field remains Gaussian.  In addition, $P(k)$ changes only in amplitude and not in shape except  in the radiation to matter  dominated  era and when there are baryon-photon interactions and baryons-dark matter interactions (see Lectures on CMB).
In detail, this is described by linear perturbation growth and by the ''transfer function".

\subsection{Examples of real world issues}
Say that now you go and try to measure a $P(k)$ from a realistic galaxy catalog. What are the real world effects you may find? We have mentioned before redshift space distortions. Here we concentrate on other effects that are more general (and not so specific to large-scale structure analysis).

\subsubsection{Discrete Fourier transform}

In the real world when you go and take the FT of your survey or even of your simulation box you will be using something like a fast Fourier transform code (FFT) which is a discrete Fourier transform.

If your box has  side of size $L$, even if $\delta(r)$ in the box is continuous, $\delta_k$ will be discrete. The k-modes sampled will be given by
\begin{equation}
\vec{k}=\left(\frac{2\pi}{L}\right)(i, j, k)\,\,\,\, {\rm where}\,\,\,\, \Delta_k=\frac{2\pi}{L}
\end{equation}

The discrete Fourier transform is obtained by placing the $\delta(x)$ on a lattice of $N^3$ grid points with spacing $L/N$. Then:
\begin{eqnarray}
\delta_k^{DFT}&=&\frac{1}{N^3}\sum_r\exp[-i\vec{k}\cdot\vec{r}]\delta(\vec{r})\\
\delta^{DFT}(\vec{r})&=&\sum_k\exp[i\vec{k}\cdot \vec{r}]\delta^{DFT}_k
\end{eqnarray}
{\bf Beware of the mapping between $r$ and $k$, some routines use a weird wrapping!}

There are different ways of placing galaxies (or particle in your simulation) on a grid: Nearest grid point, Cloud in cell, trangular shaped cloud etc...
For each of these  {\it remember}(!)  then to deconvolve the resulting $P(k)$ for their effect.
Note that 
\begin{equation}
\delta_k \sim \left( \frac{\Delta x}{2 \pi}\right)^3N^3\delta_k^{DFT}\simeq \frac{1}{\Delta k^3}\delta_k^{DFT}
\end{equation}
and thus
\begin{equation}
P(k)\simeq \frac{<|\delta^{DFT}|^2>}{(\Delta k)^3}  \,\,\,\, {\rm since} \,\,\, \delta^{D}(k)\simeq \frac{\delta^K}{(\Delta k)^3}
\end{equation}

The discretization introduces several effects: \\
The {\bf Nyquist frequency}: $k_{Ny}=\frac{2\pi}{L}\frac{N}{2}$ is  that of a mode which is sampled  by 2 grid points. Higher frequencies cannot be properly sampled and give aliasing  (spurious transfer of power) effects.  You should always work at $k< k_{Ny}$.  
There is also a minimum $k$ (largest possible scale)  that you finite box can test :$ k_{min}> 2\pi/L$. 
This is one of the --many-- reason why one needs ever larger N-body simulations...

In addition DFT assume periodic boundary conditions, if you do not have periodic boundary conditions then this also introduces aliasing.

\begin{figure}[h]
\begin{center}
\setlength{\unitlength}{1mm}
\begin{picture}(-300,120)
\includegraphics{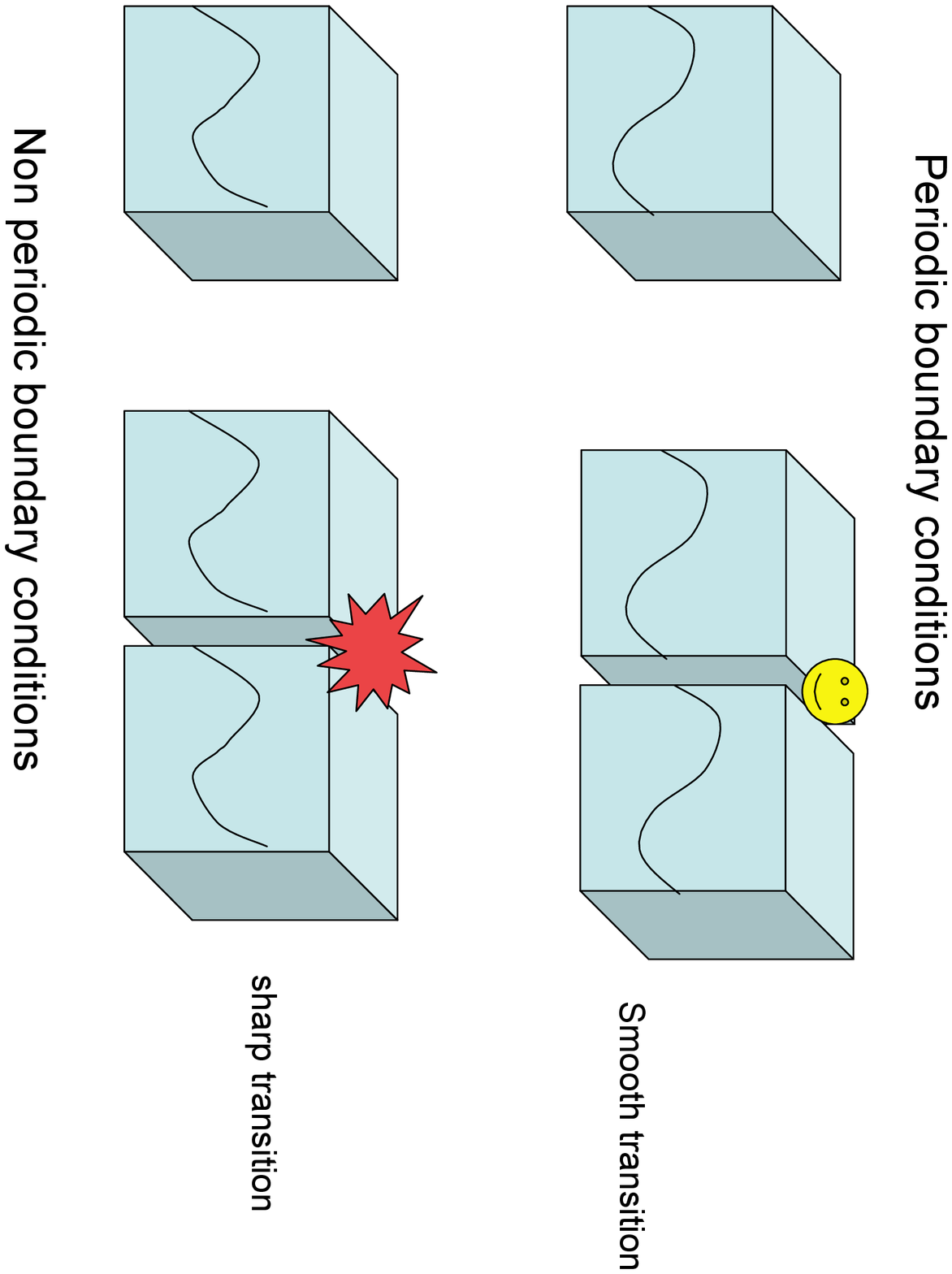}
\end{picture}
\end{center}
\caption{The importance of periodic boundary conditions}
\label{fig:2}
\end{figure}

\subsubsection{Window, selection function, masks etc}

{\bf Selection function}:  Galaxy surveys are usually magnitude limited, which means that as you look further away you start missing some galaxies.  The selection function tells you the probability for a galaxy at a given distance (or redshift $z$)  to enter the survey. It is a multiplicative effect along the line of sight in real space.

{\bf Window or  mask} You can never  observe a perfect (or even better infinite) squared box of the Universe and in CMB studies you can never have a perfect full sky map (we live in a galaxy...).  The mask  (sky cut in CMB jargon) is a function that usually takes values of $0$ or $1$and is defined on the plane of the sky (i.e. it is constant along the same  line of sight). The mask is also a real space multiplication effect. In addition sometimes in CMB studies different pixels may need to be weighted differently, and the mask is an extreme example of this where the weights are either 0 or 1.  Also this operation is a real space multiplication effect. 

Let's recall that a multiplicaton in real space (where $W(\vec{x})$ denotes the effects of window and selection functions)  
\begin{equation}
\delta^{true}(\vec{x}) \longrightarrow \delta^{obs}(\vec{x})=\delta^{true}(\vec{x}) W(\vec{x})
\end{equation}
is a convolution in Fourier space:
\begin{equation}
\delta^{true}(\vec{k}) \longrightarrow \delta^{obs}(\vec{k})=\delta^{true}(\vec{k})* W(\vec{k})
\end{equation}
the sharper $W(\vec{r})$ is the messier and delocalized $W(\vec{k})$ is.  As a result it will couple different k-modes even if the underlying ones were not correlated!
{\bf Discreteness} While the dark matter distribution is almost a continuous one the galaxy distribution is  discrete. We usually assume that the galaxy distribution is a  sampling of the dark matter distribution. The discreteness effect give the galaxy distribution a Poisson contribution (also called shot noise contribution). Note that the Poisson  contribution is non Gaussian: it is only in the limit of large number of objects (or of modes)  that it approximates a Gaussian. Here it will suffice to say that as long as a galaxy number density is high enough (which will need to be quantified and checked for any practical application) and we have enough modes,  we say that we will have a superposition of our random field (say the dark matter one characterized by its $P(k)$) plus a white noise contribution coming from the discreteness which amplitude depends on the average number density of galaxies (and should go to zero as this go to infinity), and we treat this additional contribution as if it has the same statistical properties as the underlying density field (which is an approximation).
What is the shot noise effect on the correlation properties? 

Following \cite{Peebles80} we recognize that our random field is now given by 
\begin{equation}
f(\vec{x})=n(\vec{x})=\bar{n}[1+\delta(\vec{x})]=\sum_i\delta^D(\vec{x}-\vec{x_i})
\end{equation}
where $\bar{n}$ denotes average number  of galaxies: $\bar{n}=<\sum_i\delta^D(\vec{x}-\vec{x_i})>$.
Then, as done when introducing the Poisson distribution,  we divide the volume in infinitesimal volume elements $\delta V$ so that their occupation can only be $0$ or $1$. For each of these volumes the probability of getting a galaxy is $\delta P=\rho( \vec{x})\delta V$, the probability of getting no galaxy is $\delta P= 1- \rho( \vec{x})\delta V$ and $<n_i>=<n_i^2>=\bar{n} \delta V $. We then obtain a double stochastic process with one level of randomness coming from the underlying random field and one level coming from the Poisson sampling.   
The correlation function is obtained as:
\begin{equation}
\langle \sum_{ij}\delta^D(\vec{r_1}-\vec{r_i})\delta^D(\vec{r_2}-\vec{r_j})\rangle =\bar{n}^2(1+\xi_{12})+n\delta^D(\vec{r_1}-\vec{r_2})
\end{equation}
thus
\begin{equation}
<n_1 n_2>=\bar{n}^2[1+<\delta_1\delta_2>^d] \,\,\,\, {\rm where}\,\,\, <\delta_1\delta_2>^d=\xi(x_{12})+\frac{1}{\bar{n}}\delta^D(\vec{r_1-}\vec{r_2})
\end{equation}
and in Fourier space
\begin{equation}
<\delta_{k_1}\delta_{k_2}>^d=(2\pi)^3\left(P(k)+\frac{1}{\bar{n}}\right)\delta^d(\vec{k_1}+\vec{k_2})
\end{equation}

This is not a complete surprise: the power spectrum of a superposition of two independent processes is the sum of the two power spectra....

\subsubsection{pros and cons of $\xi(r)$ and $P(k)$}
Let us briefly recap the pros and cons of working with power spectra or correlation functions.

{\bf Power spectra}\\
 {\it Pros}:  Direct connection to theory.  Modes are uncorrelated (in the ideal case). The average density ends up in $P(k=0)$ which is usually discarted, so no accurate knowledge of the mean density is needed. There is a clear distinction between linear and non-linear scales.  Smoothing is not a problem (just a multiplication)\\
 {\it Cons:} window and selection functions act as complicated convolutions, introducing mode coupling! (this is a serious issue)
 
 {\bf Correlation function}\\
 {\it Pros:} no problem with window and selection function\\
 {\it Cons:} scales are correlated. covariance calculation a real challenge even in the ideal case. Need to know mean densities very well. No clear distinction between linear and non-linear scales. No direct correspondence to theory.

 \subsubsection{... and for CMB?}
 
 If we  can observe the full sky the the CMB temperature fluctuation field can be  nicely expanded in spherical harmonics:
\begin{equation}
\Delta T(\hat{n})=\sum_{\ell>0}\sum _{m=-\ell}^{\ell} a_{\ell m}Y_{\ell m} (\hat{n}).
\end{equation}
where
\begin{equation}
a_{\ell m}=\int d\Omega_n \Delta T(\hat{n})Y_{\ell m}^* (\hat{n}).
\end{equation}
 and thus
 \begin{equation}
 <|a_{\ell m}|^2>=\langle a_{\ell m}a^*_{\ell ' m'}\rangle=\delta_{\ell \ell '} \delta_{m m'} C_{\ell}
 \end{equation}
 $C_{\ell}$ is the angular power spectrum and
 \begin{equation}
 C_{\ell}=\frac{1}{(2 \ell +1)}\sum _{m=-\ell}^{\ell}|a_{\ell m}|^2
 \end{equation}
 
 Now what happens in the presence of real world effects such as  a sky cut?
 Analogously to the real space case:
 \begin{equation}
 \widetilde{a}_{\ell m}=\int d\Omega_n \Delta T(\hat{n}) W(\hat{n}) Y^*_{\ell m}(\hat{n})
 \end{equation}
 where $W(\hat{n}) $ is a position dependent weight that in particular is set to $0$ on the sky cut.
 
 As any CMB observation gets pixelized this is 
 \begin{equation}
 \widetilde{a}_{\ell m}=\Omega_p \sum_p \Delta T(p) W(p) Y^*_{\ell m}(p)
 \end{equation}
 
 where $p$ runs over the pixels and $\Omega_p$ denotes the solid angle subtended by the pixel.
 
 Clearly this can be a problem (this can be a nasty convolution), but let us initially ignore the problem and  carry on.
 
 The pseudo-$C_{\ell}$' s (Hivon et al 2002)\cite{hivon02}  are defined as:
 \begin{equation}
 \widetilde{C}_{\ell}=\frac{1}{(2 \ell +1)}\sum _{m=-\ell}^{\ell}|\widetilde{a}_{\ell m}|^2
 \end{equation}
 
 Clearly  $\widetilde {C}_{\ell} \ne C_{\ell}$ but
 \begin{equation}
 \langle  \widetilde{C}_{\ell}\rangle= \sum_{\ell} G_{\ell \ell '} \langle  C_{\ell '}\rangle
 \end{equation}
 where $<>$ denotes the ensamble average.

We notice already two things: as expected the effect of the mask is to  couple otherwhise uncorrelated modes. In large scale structure  studies usually people stop here: convolve the theory with the various real world effects including the mask and compare that to the observed quantities. In CMB usually we go beyond this step and try to deconvolve the  real world effects.

 First of all note that
 \begin{equation}
 G_{\ell_1\ell_2}=\frac{2\ell_2+1}{4 \pi}\sum_{\ell_3} ( 2 \ell_3+1) W_{\ell_3}\left(^{l_1 l_2 l_3}_{0\,\,0\,\,0}\right)^2
 \end{equation}

 where
 \begin{equation}
 W_{\ell}=\frac{1}{2\ell+1}\sum_m |W_{\ell m}|^2 \,\,\ {\rm and} \,\,\, W_{\ell m}=\int d\Omega_n W(\hat{n})Y^*_{\ell m}(\hat{n})
 \end{equation}
 
 So if you are good enough to be able to invert $G$ and you can say that $<C_{\ell}>$ is the $C_{\ell}$ you want then 
 \begin{equation}
 C_{\ell}=\sum_{\ell '}G^{-1}_{\ell \ell '} \widetilde{C}_{\ell '}
 \end{equation}
 
Not  for all experiments it is viable (possible) to do this last step.
  
 In adddition to this, the instrument has other effects such as {\bf noise} and a finite {\bf beam}.

\subsubsection{Noise and beams}
Instrumental noise and the finite resolution of any experiment affect the measured $C_{\ell}$.  The effect of the noise is easily found: the instrumental noise is an independent  random process with a Gaussian distribution superposed to the temperature field. In $a_{lm}$ space $a_{lm}\longrightarrow a_{\ell m}^{signal}+a_{\ell m}^{noise}$. 

While $\langle a_{\ell m}^{noise}\rangle=0$, in the power spectrum this gives rise to the so-called noise bias:
\begin{equation}
C_{\ell}^{measured}=C_{\ell}^{signal}+C_{\ell}^{noise}
\end{equation}
where $C_{\ell}^{noise}=\ell (2\ell+1)\sum_m |a_{\ell m}^{noise}|^2$.  As the expectation value of $C_{\ell}^{noise}$ is non zero, this is a {\bf biased estimator}.

Note that the noise bias disappears if one computes the so-called cross $C_{\ell}$ obtained as a cross-correlation between different, uncorrelated, detectors (say detector {\it a} and {\it b}) as $\langle  a_{\ell m}^{noise,a} a_{\ell m}^{noise,b}\rangle=0$.
One is however not getting something for nothing: when one computes the covariance (or the associated error) for auto and for cross correlation $C_{\ell}$ (exercise!) the covariance is the same and includes the extra contribution of the noise. It is only that the cross-$C_{\ell}$ are {\it unbiased} estimators.

Every experiment sees the CMB with a finite resolution given by the experimental beam (similar concept to the Point Spread Function for optical astronomy). The observed temperature field is smoothed on the beam scales. Smoothing is a  convolution in real space:
\begin{equation}
T_i=\int d \Omega_n' T(\hat{n})b(|\hat{n}-\hat{n'}|)
\end{equation}
where we have considered a symmetric beam for simplicity. The beam is often well approximated by a Gaussian of a given Full Width at Half Maximum.
Remember that $\sigma_b=0.425 FWHM$.

Thus in harmonic space the beam effect is a multiplication:
\begin{equation}
C^{measured}_{\ell}=C_{\ell}^{sky}e^{-\ell^2 \sigma_b^2}
\end{equation}
and in the presence of  instrumental noise 
\begin{equation}
C^{measured}_{\ell}=C_{\ell}^{sky}e^{-\ell^2 \sigma_b^2}+C_{\ell}^{noise}
\end{equation}
 Of course, one can always deconvolve for the effects of the beam to obtain an estimate of  $C^{measured}_{\ell}$ as close as possible to $C_{\ell}^{sky}$:
 \begin{equation}
 C^{measured'}_{\ell}=C_{\ell}^{sky}+C_{\ell}^{noise}e^{\ell^2 \sigma_b^2}\,.
 \end{equation}
 
 That is why it is often said that the effective noise "blows up"  at high $\ell$ (small scales) and  why it is important to know the beam(s) well.

 ---------------------------------------------------------------------------------------
 
 {\bf Exercise}: What happens if you  use cross-$C_{\ell}$'s?
 
 {\bf Exercise}: What happens to $ C^{measured'}_{\ell}$  if  the beam is poorly reconstructed?

----------------------------------------------------------------------------------------\\ 

Note that the signal to noise of a CMB map depends on the pixel size (by smoothing the map and making larger pixels the noise per pixel will decrease as $\sqrt{\Omega_{pix}}$,  $\Omega_{pix}$ being the new pixel solid angle), on the integration time $\sigma_{pix}=s/\sqrt{t}$ where $s$ is the detector sensitivity and $t$ the time spent on a given pixel and on the number of detectors $\sigma_{pix}=s/\sqrt{M}$ where $M$ is the number f detectors.

To compare maps of different beam sizes it is useful to have a noise measure that in independent of $\Omega_{pix}$: $w=(\sigma_{pix}^2\Omega_{pix})^{-1}$.

-------------------------------------------------------------------\\

{\bf Exercise}:   Compute the expression  for  $C_{\ell}^{noise}$  given: \\
$t=$ observing time\\
$s=$ detector sensitivity (in $\mu K/\sqrt{s}$)\\
$n=$ number of detectors\\
$N=$ number of pixels \\
$f_{sky}=$ fraction of the sky observed\\
Assume uniform noise and observing time uniformly distributed.  You may find \cite{knoxthesis} very useful.

---------------------------------------------------------------\\

\subsubsection{Aside: Higher orders  correlations}
From what we have learned so far we can conclude  that the power spectrum (or the correlation function) completely characterizes  the statistical properties of the density field if it is Gaussian. But what if it is not? 

Higher order correlations are defined as: $\langle \delta_1...\delta_m \rangle$ where the deltas can be in real space giving the correlation function or in Fourier space giving power spectra.

At this stage, it  is useful to present here the  Wick's theorem (or cumulant expansion theorem). The correlation of order $m$ can in general be written as sum of products of unreducible ({\it connected})  correlations of order $\ell$ for $\ell=1...m$. For example for order 3 we obtain:
\begin{eqnarray}
&&\langle \delta_1\delta_2 \delta_3\rangle_f=\\
&&\langle \delta_1\rangle \langle \delta_2\rangle \langle \delta_3\rangle +\\
&& \langle \delta_1\rangle \langle \delta_2\delta_3\rangle+  (3 cyc. terms)\\
&&\langle \delta_1\delta_2\delta_3\rangle
\end{eqnarray}

and for order 6 (but for a distribution of zero mean):
\begin{eqnarray}
&&<\delta_1...\delta_6>_f=\\
&&<\delta_1\delta_2><\delta_3\delta_4><\delta_5\delta_6> +.. (15 terms)\\
&&<\delta_1\delta_2><\delta_3\delta_4\delta_5\delta_6>+.. (15 terms)\\
&&<\delta_1\delta_2\delta_3> <\delta_4\delta_5\delta_6>+... (10 terms)\\
&&<\delta_1..\delta_6>
\end{eqnarray}

For computing covariances of power spectra, it is useful to be familiar with  the above expansion of order 4.

 \section{More on Likelihoods}
 While the CMB temperature distribution is Gaussian (or very close to Gaussian) the $C_{\ell}$  distrIbution is not. At high  $\ell$ the Central Limit Theorem will ensure that the likelihood  is well approximated by a Gaussian but at low $\ell$ this is not the case.

\begin{equation}
{\cal L}(T|C_{\ell}^{th})\propto \frac{\exp[-(T S^{-1} T)/2]}{\sqrt{det(S)}}
\end{equation}
 where $T$ denotes a vector of the temperature map, $C_{\ell}^{th}$ denotes the  $C_{\ell}$ given by   a theoretical model (e.g. a cosmological parameters set), and $S_{ij}$ is the signal covariance:
 \begin{equation}
S_{ij}=\sum_{\ell}\frac{(2 \ell+1)}{4 \pi}C_{\ell}^{th} P_{\ell}(\hat{n_i}\cdot \hat{n_j})
 \end{equation}
 and $P_{\ell}$ denote the Legendre polynomials.
 
 If we then expand $T$ in spherical harmonics we obtain:
 \begin{equation}
 {\cal L}(T|C_{\ell}^{th})\propto \frac{\exp[-1/2 |a_{\ell m}|^2/C_{\ell}^{th}]}{\sqrt{C^{th}_{\ell}}}
 \end{equation}
 
 Isotropy means  that we can sum over $m's$ thus:
 \begin{equation}
 -2 \ln {\cal L}=\sum_{\ell}(2 \ell +1)\left[\ln \left(\frac{C^{th}_{\ell}}{C^{data}_{\ell}}\right)+\left(\frac{C_{\ell}^{data}}{C_{\ell}^{th}}\right)-1\right]
 \label{eq:likecmb1}
 \end{equation}
 where $C^{data}_{\ell}=\sum_m |a_{\ell m}|^2/(2\ell+1)$.
 
 ----------------------------------------------------------------------------\\
{\bf  Exercise}: show that for an experiment with (gaussian) noise  the expression is the same but with the substitution $C_{\ell}^{th}\longrightarrow C_{\ell}^{th}+{\cal N}_{\ell}$ with ${\cal N}$ denoting the power spectrum  of the noise.

{\bf Exercise}: show that for a partial sky experiment (that covers a fraction of sky $f_{sky}$ you can approximately write:
\begin{equation}
\ln {\cal L}\longrightarrow f_{sky} \ln {\cal L}
\end{equation}

Hint: Think about  how the number of independent modes scales with the sky area.

 --------------------------------------------------------------------------------------\\

As an aside...  you could ask: "But what do I do with polarization data?". Well... if the $a^{T}_{\ell m}$ are Gaussianly distributed also the $a^{E}_{\ell m}$  and $a^{B}_{\ell m}$  will be. So we can generalize the approach above using a vector $(a^T_{\ell m} a^E_{\ell m} a^B_{\ell m})$. Let us consider  a  full sky, ideal experiment. Start by writing down the covariance, follow the same steps as above and show that:
\begin{eqnarray}
-2\ln{\cal L}&=&\sum_{\ell} (2\ell+1)\left\{
 \ln\left(\frac{C_{\ell}^{BB}}{\hat{C}_{\ell}^{BB}}
 \right) +  \ln\left(\frac{C_{\ell}^{TT}C_{\ell}^{EE}-(C_{\ell}^{TE})^2}{\hat{C}_{\ell}^{TT}\hat{C}_{\ell}^{EE}-(\hat{C}_{\ell}^{TE})^2}\right) \right. \nonumber \\
&+&
 \frac{\hat{C}_{\ell}^{TT}C_{\ell}^{EE}+C_{\ell}^{TT}\hat{C}_{\ell}^{EE}-2\hat{C}_{\ell}^{TE}C_{\ell}^{TE}}{C_{\ell}^{TT}C_{\ell}^{EE}-(C_{\ell}^{TE})^2}+ \left. \frac{\hat{C}_{\ell}^{BB}}{C_{\ell}^{BB}}-3\right\}, \label{eq:like_ideal}
\end{eqnarray}
where $C_{\ell}$ denotes $C_{\ell}^{th}$ and $\hat{C}_{\ell}$ denotes  $C_{\ell}^{data}$. 

It is easy to show that for a noisy experiment then $C_{\ell}^{XY}\longrightarrow C_{\ell}^{XY}+{\cal N}_{\ell}^{XY}$ where $\cal{N}_{\ell}$ denotes the noise power spectrum and $X,Y=\{T, E, B\}$.

 --------------------------------------------------------------------------------------\\
 Exercise: generalize the above to partial sky coverage:  for added complication take $f_{sky}^{TT} \ne f_{sky}^{EE}\ne f_{sky}^{bb}$.   (this is often the case as the sky cut for polarization  may be different from that of temperature( the foregrounds are different) and in general the cut  (or the weighting) for $B$ may need to be larger than that for $E$.
 
 --------------------------------------------------------------------------------------\\

Following \cite{verde03} let us now expand in Taylor series Equation (\ref{eq:likecmb1}) around its maximum by writing $\hat{C}_{\ell}=C_{\ell}^{th}(1+\epsilon)$.
For a single multipole $\ell$,
\begin{equation}
-2\ln {\cal
L}_{\ell}=(2\ell+1)[\epsilon-\ln(1+\epsilon)]\simeq (2\ell+1)\left
(\frac{\epsilon^2}{2}-\frac{\epsilon^3}{3}+{\cal O}(\epsilon^4)\right)\;.
\end{equation}
We note that the Gaussian likelihood approximation is equivalent
to the above expression truncated at $\epsilon^2$: $-2 \ln {\cal L}_{{\rm Gauss},\ell}\propto(2\ell+1)/2 [(\widehat{\cal C}_{\ell}-{\cal C}_{\ell}^{\rm th})/{\cal C}_{\ell}^{\rm th}]^2\simeq (2\ell+1)\epsilon^2/2$. 

Also widely used for CMB studies is the lognormal likelihood for the equal variance approximation (Bond et al 1998):
approximation is
\begin{equation}
-2\ln {\cal L}'_{\rm LN}=\frac{(2
\ell+1)}{2}\left[\ln\left(\frac{\widehat{\cal C}_{\ell}} {{\cal
C}^{\rm th}_{\ell}}\right)\right]^2\simeq (2\ell+1)\left(\frac{\epsilon^2}{2}-
\frac{\epsilon^3}{2}\right)\;. 
\end{equation}
Thus our approximation of likelihood function is given by the form,
\begin{equation}
\ln {\cal L}=\frac{1}{3} \ln {\cal L}_{\rm Gauss}+\frac{2}{3}\ln {\cal L}'_{\rm LN}\;,
\label{eq:likelihoodform}
\end{equation}
where 

\begin{equation}
\ln {\cal L}_{\rm Gauss}\propto -\frac{1}{2}\sum_{\ell\ell'}
({\cal C}_{\ell}^{\rm th}-\widehat{\cal C}_{\ell})Q_{\ell \ell'}
({\cal C}_{\ell'}^{\rm th}-\widehat{\cal C}_{\ell'})\;,
\label{eq:like_gauss}
\end{equation}

and 
\begin{equation}
\ln {\cal L}_{\rm LN}=-1/2\sum_{\ell
\ell'}(z_{\ell}^{\rm th}-\widehat{z}_{\ell}){\cal Q}_{\ell
\ell'}(z_{\ell'}^{\rm th}-\widehat{z}_{\ell'}),
\label{eq:lognormallikelihood}
\end{equation}
\noindent where $z_{\ell}^{\rm th}=\ln({\cal C}_{\ell}^{\rm th}+{\cal N}_{\ell})$, 
$\widehat{z}_{\ell}=\ln(\widehat{\cal C}_{\ell}+{\cal N}_{\ell})$  and ${\cal Q}_{\ell \ell'}$ is the local transformation of the curvature matrix $Q$ to the lognormal variables $z_{\ell}$, 
\begin{equation}
{\cal Q}_{\ell \ell'}=({\cal C}^{th}_{\ell}+{\cal N}_{\ell})Q_{\ell \ell'}(\widehat{\cal C}^{th}_{\ell'}+{\cal N}_{\ell'}).
\label{eq:lognormalvariables}
\end{equation}
 The curvature matrix is the inverse of the covariance matrix evaluated at the maximum likelihood. However we do not want to adopt the ''equal variance approximation", so $Q$ for us will be in inverse of the covariance matrix.
 
 The elements of the covariance matrix, even for a ideal full sky experiment can be written as:
 \begin{equation}
{\bf C}_{\ell \ell}= 2\frac{({\cal C}_{\ell}^{th})^2}{2\ell+1}
 \end{equation}
 
 In the absence of noise the covariance is non zero: this is the {\bf cosmic variance}.

 Note that for the latest WMAP release \cite{spergel07, page07}at low $\ell $ the likelihood is computed directly from the maps $\vec{m}$. 
The standard likelihood is given by
\begin{equation}
 L(\vec{m}|S)d\vec{m} = 
\frac{\exp\left[-\frac12\vec{m}^t(S+N)^{-1}\vec{m}\right]}{|S+N|^{1/2}}
\frac{d\vec{m}}{(2\pi)^{3n_p/2}},
\label{eq:like}
\end{equation}
where $\vec{m}$ is the data
vector containing the temperature map, $\vec{T}$, as well as the 
polarization maps, $\vec{Q}$, and $\vec{U}$, 
$n_p$ is the number of pixels of each map, and $S$ and $N$ are
the signal and noise covariance matrix ($3n_p\times 3n_p$), 
respectively.
As the temperature data are completely dominated
by the signal at such low multipoles, noise in temperature 
may be ignored. This simplifies the form of likelihood as
\begin{equation}
 L(\vec{m}|S)d\vec{m} =
\frac{\exp\left[-\frac12\vec{\tilde{m}}^t(\tilde{S}_P+N_P)^{-1}
\vec{\tilde{m}}\right]}{|\tilde{S}_P+N_P|^{1/2}}
\frac{d\vec{\tilde{m}}}{(2\pi)^{n_p}}~ 
\frac{\exp\left(-\frac12\vec{T}^tS_{T}^{-1}\vec{T}\right)}{|S_{T}|^{1/2}}
\frac{d\vec{T}}{(2\pi)^{n_p/2}},
\label{eq:dnslike}
\end{equation}
where $S_T$ is the temperature signal matrix ($n_p\times n_p$),
the new polarization data vector, 
$\vec{\tilde{m}}=(\tilde{Q}_p,~\tilde{U}_p)$ and $\tilde{S}_P$ is the signal matrix for the new polarization vector
with the size of $2n_p\times 2n_p$. 

At the time of writing,  in CMB  parameter estimates for $\ell <~ 2000$, the likelihood calculation is the bottleneck of the analysis.
 
 \section{Monte Carlo methods}
 \subsection{Monte Carlo error estimation}
  Let's go back to the issue of  parameter estimation and error calculation.
 Here is the conceptual interpretation of what it means that en  experiment measures some parameters (say cosmological parameters). There is some underlying true set of parameters $\vec{\alpha}_{true}$ that are only known to Mother Nature but  not to the experimenter. There true parameters are statistically realized in the observable universe and random measurement errors are then included when the observable universe gets measured. This ''realization" gives the measured data ${\cal D}_0$. Only ${\cal D}_0$ is accessible to the observer (you). Then you go and do what you have to do to estimate the parameters and their errors (chi-square, likelihood,  etc...) and get $\vec{\alpha}_0$. Note that ${\cal D}_0$ is not a unique  realization of the true model given by $\vec{\alpha}_{true}$ : there could be  infinitely many other realizations as {\it hypothetical data sets}, which could have been the measured one:  ${\cal D}_1, {\cal D}_2, ...$ each of them with a slightly different fitted parameters $\vec{\alpha}_1, \vec{\alpha}_2...$.  $\vec{\alpha}_0$ is one parameter set drawn from this distribution. The hypotetical ensamble of universes  described by  $\vec{\alpha}_i$ is called ensamble, and one expects that the expectation value $\langle \vec{\alpha}_i\rangle=\vec{\alpha}_{true}$.
 If we knew the distribution of $\vec{\alpha}_i-\vec{\alpha}_{true}$ we would know everything we need about the uncertainties in our measurement $\vec{\alpha}_0$. The goal is to infer the distribution of $\vec{\alpha}_i-\vec{\alpha}_{true}$ without knowing $\vec{\alpha}_{true}$.

Here's what we do: we say that hopefully $\vec{\alpha}_0$ is not too wrong and we consider a fictitious world where $\vec{\alpha}_0$ was the true one. So it would not be such a big mistake to take  the probability distribution of $\vec{\alpha}_0-\vec{\alpha}_i$ to be that of $\vec{\alpha}_{true}-\vec{\alpha}_i$. In many cases we know how to simulate $\vec{\alpha}_0-\vec{\alpha}_i$ and so we can simulate many synthetic realization of "worlds" where $\vec{\alpha}_0$ is the true underlying model. Then mimic the observation process of these fictitious Universes replicating all the observational errors and effects and from each of these fictitious universe estimate the parameters. Simulate enough of them and  from $\vec{\alpha}^S_i-\vec{\alpha}_0$ you will be able to map the desired multi-dimensional probability distribution.
 
 With the advent of fast computers this technique has become increasingly widespread. As long as you believe you know the underlying distribution and that you believe you can mimic the observation replicating all the observational  effects this technique is extremely powerful and, I would say, indispensable.

 \subsection{Monte Carlo Markov Chains}
 
  When dealing with high dimensional likelihoods (i.e. many parameters) the process of mapping the likelihood  (or the posterior) surface  can become very expensive. For example for CMB studies the modles considered have from 6 to 11+ parameters.  Every model evaluation even with a fact code such as CAMB can take up to minutes per iteration. A grid-based likelihood analysis would require prohibitive amounts of CPU
time. For example, a coarse grid ($\sim 20$ grid points 
per dimension) with six parameters requires 
$\sim 6.4\times 10^7$ evaluations of the power spectra. 
At 1.6 seconds per evaluation, the calculation would take $\sim 1200$
days.  Christensen \& Meyer (2000) \cite{christensenmeyer} proposed using Markov Chain Monte Carlo 
(MCMC) to investigate the likelihood space.  This approach has become the standard 
tool for CMB analyses. MCMC is a method to simulate posterior distributions. In particular one simulates sampling the  posterior distribution ${\cal P}({\bf \alpha}|x)$, of a set of parameters ${\bf \alpha}$ given event $x$, obtained via Bayes' Theorem %
\begin{equation}
{\cal P}(\alpha|x)=\frac{{\cal P}(x|\alpha){\cal P}(\alpha)}{\int
{\cal P}(x|\alpha){\cal P}(\alpha)d\alpha},
\label{eq:bayes2}
\end{equation}
\noindent where ${\cal P}(x|\alpha)$ is the likelihood of event $x$ given the model parameters $\alpha$ and ${\cal P}(\alpha)$ is the prior probability density; $\alpha$ denotes a set of cosmological parameters (e.g., for the standard, flat $\Lambda$CDM model these could be, the cold-dark matter density parameter $\Omega_c$, the baryon density parameter $\Omega_b$, the spectral slope $n_s$, the Hubble constant --in units of $100$ km s$^{-1}$ Mpc$^{-1}$)-- $h$, the optical depth $\tau$ and the power spectrum amplitude $A$), and event $x$ will be the set of observed $\widehat{\cal C}_{\ell}$. The MCMC generates random draws (i.e. simulations) from the posterior distribution that are a ``fair'' sample of the likelihood surface. From this sample, we can estimate all of the quantities of interest about the posterior distribution (mean, variance, confidence levels). The MCMC method scales approximately linearly with the number of parameters, thus allowing us to perform likelihood analysis in a reasonable amount of time.

A properly derived and implemented MCMC draws from the joint posterior density ${\cal P}(\alpha|x)$ once it has converged to the stationary distribution.  The primary consideration in implementing MCMC is determining when the chain has  {\it converged}. After an initial {\it ``burn-in''} period, all further samples can be thought of as coming from the stationary distribution. In other words the chain has no dependence on the starting location. 
 
\begin{figure}
\begin{center}
\setlength{\unitlength}{1mm}
\begin{picture}(-300,100)
\includegraphics{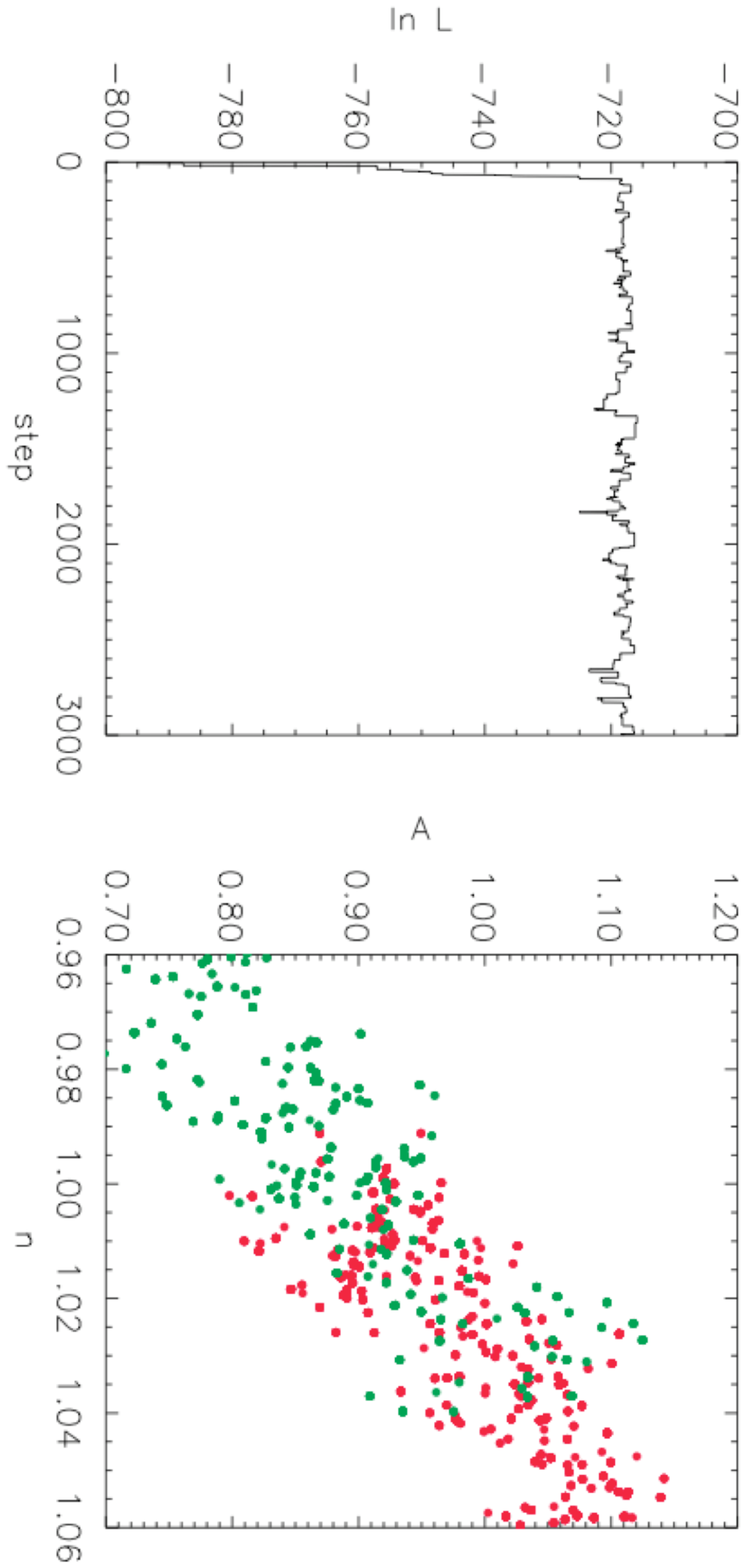}
\end{picture}
\end{center}
\caption{Unconverged Markov chains. The left panel shows a {\bf trace plot} of the likelihood values versus  iteration number for one MCMC (these are the first 3000 steps from  run). Note 
the burn-in for the first 100 steps. In the right panel, red dots are points of the chain in the (n, A) plane 
after discarding the burn-in. Green dots are from another MCMC for the same data-set and the same model. 
It is clear that, although the trace plot may appear to indicate that the chain has converged, it has not fully 
explored the likelihood surface. Using either of these two chains at this stage will give incorrect results for 
the best fit cosmological parameters and their errors. Figure from ref \cite{verde03}}
\label{fig:unconv}
\end{figure}

Another fundamental problem of inference from Markov chains is that there are always  areas of the target distribution that have not been covered by a finite chain. If the MCMC is run for a very long time, the ergodicity of the Markov chain guarantees that eventually the chain will cover all the target distribution, but in the short term the simulations cannot tell us about areas where they have not been. It is thus crucial that the chain achieves good {\it ``mixing''}. If the
Markov chain does not move rapidly throughout the support of the target distribution because of poor {\it mixing}, it might take a prohibitive amount of time for the chain to fully explore the likelihood surface.  Thus it is important to have a convergence criterion and a mixing diagnostic.  Plots of the sampled MCMC parameters or likelihood values versus iteration number are commonly used to provide such criteria (left panel of Figure \ref{fig:unconv}). However, samples from a chain are typically serially correlated; very high auto-correlation leads to little movement of the chain and thus makes the chain to ``appear'' to have converged. For a more detailed discussion see \cite{gilk}. Using a MCMC that has not fully explored the likelihood surface for determining cosmological parameters will yield {\it wrong} results. See right  panel of  Figure \ref{fig:unconv}).

\subsection{Markov Chains in Practice}
 Here are the necessary steps to run a simple MCMC for the CMB temperature power spectrum. It is straightforward to generalize these instructions to include the temperature-polarization power spectrum and other datasets. 
The MCMC is essentially  a random walk in parameter space, where the probability of being at any position in the space is proportional to the posterior probability.
\begin{itemize}
\item[1)] Start with a set of cosmological parameters $\{\alpha_1\}$, compute the ${\cal C}^{1}_{\ell}$ and the likelihood ${\cal L}_1={\cal L}({\cal C}^{1 {\rm th}}_{\ell}|\widehat{\cal C}_{\ell})$. 
\item[2)] Take a random step in parameter space to obtain a new set of cosmological parameters $\{\alpha_2\}$. The probability distribution of the step is taken to be Gaussian in each direction $i$ with r.m.s given by $\sigma_i$. We will refer below to $\sigma_i$ as the ``step size''. The choice of the step size is important to optimize the chain efficiency (see discussion below)
\item[3)] Compute the ${\cal C}^{2 {\rm th}}_{\ell}$ for the new set of cosmological parameters and their likelihood ${\cal L}_2$.
\item[4.a)] If ${\cal L}_2/{\cal L}_1 \ge 1$, ``take the step'' i.e. save the new set of cosmological parameters $\{\alpha_2\}$ as part of the chain, then go to step 2 after the substitution $\{\alpha_1\}\longrightarrow \{\alpha_2\}$.
\item[4.b)] If ${\cal L}_2/{\cal L}_1 < 1$, draw a random number $x$ from a uniform distribution from 0 to 1. If $x \ge {\cal L}_2/{\cal L}_1 $ ``do not take the step'', i.e. save the parameter set $\{\alpha_1\}$ as part of the chain and return to step 2. If $x < {\cal L}_2/{\cal L}_1 $, `` take the step'', i.e. do as in 4.a).
\item[5)]For each cosmological model run four chains starting at randomly chosen, well-separated points in parameter space.  When the convergence criterion is satisfied and the chains have enough points to provide reasonable samples from the {\rm a posteriori} distributions (i.e. enough points to be able to reconstruct the 1- and 2-$\sigma$ levels of the marginalized likelihood for all the parameters) stop the chains. 
\end{itemize}
It is clear that the MCMC approach is easily generalized to compute the joint likelihood of {\sl WMAP} data with other datasets.

\subsection{Improving MCMC Efficiency}
MCMC efficiency can be seriously compromised if there are degeneracies among parameters.  The typical example is the degeneracy between $\Omega_m$ and $H$  for a flat cosmology or that between $\Omega_m$ and $\Omega_{\Lambda}$ for a non-flat case (see e.g. reference \cite{spergel07}).

The Markov chain efficiency can be improved in different ways.  Here we report the simplest way.

 {\bf Reparameterization}

We describe below the method we use to ensure convergence and good mixing.
Degeneracies and poor parameter choices slow the rate of convergence and mixing of the Markov Chain. There is one near-exact degeneracy (the geometric degeneracy) and several approximate degeneracies in the parameters describing the CMB power spectrum  Bond et al (1994) \cite{bond94}, Efstathiou\& Bond(1984) \cite{be84}. 
The numerical effects of these degeneracies are reduced by finding a combination of cosmological parameters (e.g., $\Omega_c$, $\Omega_b$, $h$, etc.) that have essentially orthogonal effects on the angular power spectrum. The use of such parameter combinations removes or reduces degeneracies in the MCMC and hence speeds up convergence and improves mixing, because the chain does not have to spend time exploring degeneracy directions. Kosowsky, Milosavljevic \& Jimenez (2002) \cite{kmj02}  and Jimenez et al (2003) \cite{cmbwarp} introduced a set of reparameterizations to do  just this. In addition, these new parameters  reflect the underlying physical effects determining the form of the CMB power spectrum (we will refer to these as physical parameters). This leads to particularly intuitive and transparent parameter dependencies of the CMB power spectrum.

For the 6 parameters LCDM model these ''normal" or ''physical" parameters are: the physical energy densities of cold dark matter, $\omega_c\equiv \Omega_c
h^2$, and baryons, $\omega_b\equiv \Omega_b h^2$, the characteristic angular scale of the acoustic peaks,
\begin{equation}
\theta_A = \frac{r_s(a_{dec})}{D_A(a_{dec})},
\end{equation} 
\noindent where $a_{dec}$ is the scale factor at decoupling,
\begin{eqnarray}
r_s(a_{dec})&=&\frac{c}{H_0\sqrt{3}} \times \\ \nonumber 
&& \int_0^{a_{dec}} \frac{dx}{
\left[\left(1+\frac{3\Omega_b}{4\Omega_\gamma}\right)\left((1-\Omega)x^2+\Omega_\Lambda
x^{1-3w}+\Omega_m x + \Omega_{rad}\right)\right]^{1/2}}
\end{eqnarray}
is the sound horizon at decoupling, and 
\begin{equation}
d_A(a_{dec})=\frac{a}{H_0}\frac{S_{\kappa}(r_{dec})}{\sqrt{|\Omega-1|}}
\end{equation}
where
\begin{equation}
r(a_{dec})=|\Omega-1|\int_{a_{dec}}^1\frac{dx}{\left[(1-\Omega)x^2+\Omega_\Lambda x^{1-3w}+\Omega_m x + \Omega_{rad}\right]^{-1/2}}
\end{equation}

and $S_{\kappa}(r)$ as usual coincides with the argument if the curvature $\kappa$ is $0$,  is a $\sin$ function for $\Omega>1$ and a $\sinh$ function otherwhise.
Here $H_0$ denotes the Hubble constant and $c$ is the speed of light,
$\Omega_m=\Omega_c+\Omega_b$, $\Omega_{\Lambda}$ denotes  the dark energy density parameters, $w$ is the equation of state of the dark energy component,  $\Omega=\Omega_m+\Omega_{\Lambda}$ and  the radiation density parameter
$\Omega_{\rm rad}=\Omega_{\gamma}+\Omega_{\nu}$, $\Omega_{\gamma}$, $\Omega_{\nu}$ are the the photon and neutrino density parameters respectively. 
For reionization sometmes the parameter  ${\cal Z}\equiv \exp(-2\tau)$  is used, where $\tau$ denotes the 
optical depth to the last scattering surface (not the decoupling
surface). 

These reparameterizations are useful because the degeneracies  are non-linear, that is they are not well described by ellipses in parameter space.  For degeneracies that are well approximated by ellipses in parameter space it is possible to find the best reparameterization automatically. This is what the code {\bf CosmoMC} \cite{cosmomc, cosmomcpaper} (see tutorials) does.  To be more precise it computes the  parameters covariance matrix from which the axes of the multi-dimensional degeneracy ellipse  can be found. Then it performs a rotation and re-scaling of the coordinates (i.e. the parameters) to transform the  degeneracy ellipse  in an azimutally symmetric contour. See  discussion at \verb1 http://cosmologist.info/notes/CosmoMC.pdf 1 for more information. This  technique can improve the MCMC efficiency  up to a factor of  order 10.

{\bf Step size optimization} The choice of the step size in the Markov Chain is crucial to improve the chain efficiency and speed up convergence. If the step size is too big, the acceptance rate will be very small; if the step size is too small the acceptance rate will be high but the chain will exhibit poor mixing. Both situations will lead to slow convergence.

\subsection{Convergence and Mixing}
\label{sec.conv}
Before we even start this section: 
{\bf thou shall always use a convergence and mixing criterion when running MCMC's}.

Let's illustrate here the method proposed by Gelman \& Rubin 1992 \cite{converge} as an example.
They advocate comparing several sequences drawn from different 
starting points and checking to see that they are indistinguishable. 
This method not only tests convergence but can also diagnose poor
mixing. Let us consider $M$ chains  starting at well-separated points in parameter space; each has $2N$ elements, of which we consider only the last N: $\{y_i^j\}$ where $i=1,..,N$ and $j=1,..,M$, i.e. $y$ denotes a chain element (a point in parameter space) the index $i$ runs over the elements in a chain the index $j$ runs over the different chains. We define the mean of the chain
\begin{equation}
\bar{y}^j=\frac{1}{N}\sum_{i=1}^{N}y_i^j,
\end{equation}
and the mean of the distribution
\begin{equation}
\bar{y}=\frac{1}{NM}\sum_{ij=1}^{NM}y_i^j\,.
\end{equation}
We then define the variance between chains as
\begin{equation}
B_n=\frac{1}{M-1}\sum_{j=1}^M(\bar{y}^j-\bar{y})^2,
\end{equation}
and the variance within a chain as
\begin{equation}
W=\frac{1}{M(N-1)}\sum_{ij}(y^j_i-\bar{y}^j)^2.
\end{equation}
The quantity
\begin{equation}
\hat{R}=\frac{\frac{N-1}{N}W+B_n\left(1+\frac{1}{M}\right)}{W}
\end{equation}
is the ratio of two estimates of the variance in the target distribution: the numerator is an estimate  of the variance that is unbiased if the distribution is stationary, but is otherwise an overestimate. The denominator is an underestimate of the variance of the target distribution if the individual sequences did not have time to converge.

The convergence of the Markov chain is then monitored by recording the quantity $\hat{R}$ for all the  parameters and running the simulations until the values for $\hat{R}$ are always $< 1.03$.
Needless to say that the cosmomc package offers several convergence and mixing  diagnostic tools as part of the ``getdist'' routine.

---------------------------------------------------------------------------------------------

Question: how does the MCMC sample the prior if all one actually computes is the likelihood?

-----------------------------------------------------------------------------------------------

\subsection{MCMC Output Analysis}
Now that you have your multiple chains  and the convergence criterium says they are converged whot do you do?
 First discard {\it  burn in}  and merge the chains. Since the MCMC passes objective tests for convergence and mixing, the density of points in parameter space is proportional to the posterior probability of the parameters.  (Note that cosmomc saves repeated steps as the same entry in the file but with a weight equal to the repetitions: the MCMC gives to each point in parameter space a ``weight'' proportional to the number of steps the chain has spent at that particular location.).
The marginalized distribution is obtained by projecting the MCMC points. This is a great advantage compared to the grid-based approach where multi-dimensional integrals would have to be performed. The MCMC basically performs a Monte Carlo integration. 
the density of points in the n-dimensional space is proportional to the posterior, and best fit parameters and  multi-dimensional confidence levels can be found as illustrated in the last class.

Note that  the global maximum likelihood value for the parameters does not necessarily coincide with the expectation value of their marginalized distribution if the likelihood surface is not a multi-variate Gaussian.

A virtue  of the MCMC method is that the addition of extra data sets in the joint analysis can efficiently be done with minimal computational effort from the MCMC output if the inclusion of extra data set does not require the introduction of extra parameters or  does not drive the parameters significantly away from the current best fit. 
If the likelihood surface for a subset of parameters from an external (independent) data set is known, or if a prior needs to be added {\it a posteriori}, the joint posterior surface can be obtained by multiplying the new probability distribution with the posterior distribution of the MCMC output. 
 To be more precise:  as the density of point (i.e. the weight) is directly proportional to the posterior, then this is achieved by multiplying the weight by the new probability distribution.

The cosmomc package already includes this facility.
 
 \section{Fisher Matrix}
 
 What if you wanted  to forecast how well a future experiment can do?
There is the  expensive but more accurate way and the cheap and quick way (but often less accurate).
The expensive way is in the same spirit of Monte-Carlos simulations discussed earlier: simulate the observations and estimate the parameters as you would do on the data. However often you want to have a much quicker way to forecasts parameters uncertainties, especially  if you need to quickly compare many different experimental designs. This technique is the Fisher matrix approach.

\section{Fisher matrix}
The question of how accurately one can measure model parameters from a given data set (without simulating the data set) was answered more than  70 years ago by Fisher (1935) \cite{Fisher}.
Suppose your data set is given by $m$ real numbers $x_1...x_m$ arranged in a vector (they can be CMB map pixels, $P(k)$ of galaxies etc...) $\vec{x}$ is a random variable with probability distribution  which depends in some way on the vector of model parameters $\vec{\alpha}$.
The Fisher information matrix id defined as:
\begin{equation}
F_{ij}=\left \langle  \frac{\partial^2 L}{\partial \alpha_i \partial \alpha_j}\right \rangle 
\label{eq:fisher}
\end{equation} 
where $L=-\ln {\cal L}$.
In practice  you will choose a fiducial model and compute the above at the fiducial model. 
In the one parameter   case let's  note that if the Likelihood is Gaussian then $L=1/2 (\alpha-\alpha_0)/\sigma_{\alpha}^2$
where $\alpha_0$ is the value that maximizes the likelihood and sigma is the error on the parameter $\alpha$. Thus the second derivative wrt  $\alpha$ of $L$ is $1/\sigma_{\alpha}^2$ as long as $\sigma_{\alpha}$ does not depend on $\alpha$. In general we can expand $L$ in Taylor series around its maximum (or the fiducial  model). There by definition the first derivative of $L$ wrt the parmeters is 0.
 
 \begin{equation}
 \Delta L=\frac{1}{2}\frac{d^2L}{d\alpha^2} (\alpha-\alpha_0)^2
 \end{equation}

When  $2\Delta L=1$, which in the case when we can identify it with  $\Delta \chi^2$ with  corresponds to 68\% (or 1-$\sigma$) then $1/\sqrt{d^2L/d\alpha^2}$ is the 1 sigma displacement of $\alpha$ from $\alpha_0$.

The generalization to many variables is beyond our scope here  (see Kendall \& Stuard 1977 \cite{kendallstewardt77}) let's just say that an estimate of the covariance for the parameters is given by:
 \begin{equation}
 \sigma^2_{\alpha_i, \alpha_j}\ge ({\bf F^{-1}})_{ij}
 \end{equation}

From here it follows that if all other parameters are kept fixed
\begin{equation}
\sigma_{\alpha_i}\ge \sqrt{\frac{1}{F_{ii}}}
\end{equation}
(i.e. the reciprocal of the square root of the diagonal element $ii$ of the Fisher matrix.)

But if all other parameters are estimated from the data as well then the marginalized error is 
\begin{equation}
\sigma_{\alpha}=({\bf F^{-1}})_{ii}^{1/2}
\end{equation}
(i.e. square root of the element $ii$ of the inverse of the Fisher matrix -perform a matrix inversion here!)

 In general, say that  you have 5 parameters and that you want to plot the joint 2D contours for parameters 2 and 4 marginalized over all other parameters 1,3,5.
Then you invert $F_{ij}$, take the minor 22, 24, 42, 24  and invert it back. The resulting matrix, let's call it Q, describes a Gaussian 2D likelihood surface in the parameters  2 and 4  or, in other words,   the chisquare  surface for parameters 2,4 - marginalized over all other parameters- can be described by the equation
$\tilde{\chi}^2=\sum_{kq} (\alpha_k-\alpha_k^{fiducial}) Q_{kq} (\alpha_q-\alpha_q^{fiducial})$.

From  this equation,  getting the errors corresponds to finding the quadraitic equation solution  $\tilde{\chi}^2=\Delta \chi^2$. 
For correspondence between $\Delta \chi^2$ and confidence region see  the earlier discussion.
If you want ti make plots, the equation for the elliptical boundary  for the joint confidence region  in the sub-space of parameters of interest is:
$\Delta =\delta \vec{\alpha} Q^{-1} \vec{\delta \alpha}$.

Note that in many applications the likelihood for the 
data is assumed to be Gaussian and the data-covariance is assumed not 
to depend on the parameters. For example for the application to CMB the fisher matrix is often computed as:
\begin{equation}
F_{ij}=\sum_{\ell} \frac{(2 \ell+1)}{2}\frac{\frac{\partial C_{\ell}}{\partial \alpha_j}\frac{\partial C_{\ell}}{\partial \alpha_j}}{(C_{\ell}+{\cal N} e^{\sigma^2 \ell^2})^2}
\end{equation}
This approximation is good in the high $\ell$, noise dominated regime. In this regime the central limit theorem ensures a Gaussian likelihood and  the cosmic variance contribution to the covariance is negligible and thus the covariance does not depend on the cosmological parameters. However at low $\ell$ in the cosmic variance dominated regime,  this approximation over-estimates the errors (this is one of the few cases where the Fisher matrix approach over-estimates errors) by about a factor of 2.  In this case it is therefore preferable to go for the more numerically intensive option of computing  eq. \ref{eq:fisher} with the exact form for the likelihood Eq \ref{eq:like_ideal}.

 Before we conclude we report here a  useful identity:
 
 \begin{equation}
 2 L=\ln det(C)+(x-y)_iC^{-1}_{ij}(x-y)_j^T=Tr[\ln C+C^{-1}D]
 \end{equation}

where $C$ stands for the covariance matrix of the data, repeated indices are summed over, $x$ denotes the data and $y$ the model  fitting the data and $D$ is the data matrix defined as $(\vec{x}-\vec{y})(\vec{x}-\vec{y})^t$. We have used the identity $\ln det(C)=Tr \ln C$.

\section{Conclusions}

Wether you are a theorist or an experimentalist in cosmology, these days you cannot ignore the fact that  to make the most of your data, statistical techniques need to be employed, and used correctly. An incorrect treatment of the data will lead to nonsensical results.  A given data set can revel a lot about the universe, but there will always things that are beyond the statistical power of the data set. It is crucial to recognize it. I hope I have given you the basis to be able to learn this by yourself.  When in doubt  always remember:   treat your data with respect.

\section*{Acknowledgments}

We would like to thank Alain Connes, Jeff Harvey,
Albert Schwarz and Nathan Seiberg for
comments on the manuscript, and the ITP, Santa Barbara for hospitality.
This work was supported in part by DOE grant DE-FG05-90ER40559, and by
RFFI grants 01-01-00549 and 00-15-96557.
 \section{Questions}
 Here are some of the questions (and their answers...)
 
Q:  What about the forgotten ${\cal P}(D)$ in the Bayes theorem?\\
A:  ${\cal P}(D)$ becomes important when one wants to compare different cosmological models (not just different parameter values  within the same cosmological model). There is a somewhat extensive literature in cosmology alone on this: ``Bayesian evidence'' is used to do this ``model selection''. Bayesian evidence calculation can be, in many cases, numerically intensive.

Q: What do you do when  the likelihood surface is very complicated for example is multi-peaked?\\
A:  Luckily enough in CMB analysis this is almost never the case (exceptions being, for example, the cases where sharp oscillations are present in the   $C^{th}_{\ell}$ as it happens in transplanckian models.) In other contexts this case is more frequent. In these cases, when additional peaks can't be suppressed by a motivated prior, there are several options: {\it a)} if $m$ is the number of parameters,  report the confidence levels by considering only regions with $m$-dimensional likelihoods above a threshold before marginalizing (thus local maxima will be included in the confidence contours if significant enough) or  {\it b)} simply marginalize as described here, but expect that the marginalized peaks will not coincide with the $m$-dimensional peaks. The most challenging issue when the likelihood surface is complicated is having the MCMC to fully explore the surface. Techniques such as Hamiltonian Monte Carlo are very powerful in these cases see e.g. \cite{hmc1, hmc2}

Q: The  Fisher matrix approach is very interesting,  is there anything I should look out for? \\
A:  The Fisher matrix approach assumes that the parameters log-likelihood surface can be quadratically approximated around the maximum. This may or may not be a good approximation. It is always a good idea to check at least in a reference case wether this approach significantly underestimated the errors. 
In many Fisher matrix implementation the likelihood for the data is assumed to be Gaussian and the data-covariance is assumed not to depend on the parameters. While this approximation  greatly simplifies the calculations ({\bf Exercise:} show why this is the case), it may significantly mis-estimate the size of the errors. 
 In addition, the  Fisher matrix  calculation often requires numerical evaluation of second derivatives. Numerical derivatives always need a lot of care and attention: you have been warned.
 
Q:  Does  cosmomc  use the sampling described here?\\
A:  The recipe reported here is the so called Metropolis-Hasting algorithm. Cosmomc  offers also other sampling algorithms: slice sampling,  or a split between ``slow" and ``fast'' parameters or the learn-propose option where chains automatically adjust the proposal distribution  by repeatedly computing the parameters covariance matrix on the fly. These techniques  can greatly improve the MCMC  ''efficiency". See for example \\
\verb1 http://cosmologist.info/notes/CosmoMC.pdf1 or \cite{gilk} for more details.

\subsubsection*{Acknowledgments}
I am indebited to A. Taylor  for his lectures on statistics for beginners given at ROE in 1997. Also a lot of my understanding of the basics come from  Ned Wright ''Journal Club in statistics" on his  web page.
A lot  of the techniques reported here were developed and/or tested as part of the analysis of WMAP data between 2002 and 2006. Thanks also to E. Linder and Y. Wang for reading and sending comments on the manuscript. Last, but not least, I would like to thank the organizers of the XIX Canary islands winter school, for a  very stimulating school.

\section{Tutorials} 
 Two tutorial sessions were organized as part of the Winter School. You may want to try to repeat the same steps.
 
  The goals of  the first tutorial  were:
 \begin{itemize}
\item Download and install Healpix \cite{healpixpaper, healpixweb}. make sure you have a  fortran 90 compiler installed and possibly also IDL installed
\item The LAMBDA site \cite{LAMBDA} contains a lot of extremely useful information: browse the page.
\item Find the on-line calculator for the theory $C_{\ell}$ given some cosmological parameters, generate a set of $C_{\ell}$ and save them as a "fits" file.
\item Browse the help pages of Healpix  and find out what it does.
\item In particular the routine ''symfast" enables one  to generate a map from a set of theory $C_{\ell}$.
\item The routine "anafast" enables one to compute $C_{\ell}$ from a map.
\item  Using "symfast", generate two maps with two different random number generators for the $C_{\ell}$ you generated above. Select a beam of $FWHM$ of, say, $30'$, but for now do not imose any galaxy cut and do not add noise. Compare the maps. To do that use "mollview". Why are they different?
\item  Using "anafast" compute the $C_{\ell}$  for both maps.
\item Using your favorite  plotting routine (the IDL part of healpix offers utilities to do that) plot the orginal $C_{\ell}$ and the two realizations.  Why do they differ?
\item Deconvolve for the beam, and make sure you are plotting the correct units, the correct factors of $\ell (\ell+1)$  etc.  Why do the power spectra still differ?
\item Try to compute the $C_{\ell}$ for $\ell_{max}=3000$ and for a non flat model with camb. It takes much longer than for a simple, flat LCDM model. A code like CMBwarp \cite{cmbwarp} offers a shortcut. try it by downloading it at \verb1 http://www.astro.princeton.edu/~raulj/CMBwarp/index.html1. Keep in mind however that this offers a fast fitting routine for a fine grid of  pre-computed $C_{\ell}$, it is not a Boltzmann  code and thus is valid only within the  quoted limits!
\end{itemize}

The goals of  the second  tutorial  were:
\begin{itemize}
\item Download and install the CosmoMC \cite{cosmomc} package (and read the instructions).
\item Remember that WMAP data and likelihood need to be dowloaded  separately from the LAMBDA site. Do this following the instructions.
\item  Take a look at the params.ini file. In here you set up the MCMC
\item   Set a chain to run.
\item get familar with the getdist program and distparams.ini file. This program checks convergence for you and compute basic parameter estimation from converged chains.
\item download from LAMBDA a set of chains, the suggested one was for the flat, quintessence model for WMAP data only.
\item  plot the marginalized probability distribution for the parameter $w$.
\item {\bf the challenge}:  Apply now a  Hubble constant prior to this chain, take the HST key project \cite{hstkey} constraint of $H_0= 72 \pm 8$ km/s/Mpc and assume it has a Gaussian distribution.   This procedure  is called ''importance sampling".
\end{itemize}

\bibliographystyle{apsrmp}


\begin{thebibliography}{99}

\expandafter\ifx\csname natexlab\endcsname\relax\def\natexlab#1{#1}\fi
\expandafter\ifx\csname bibnamefont\endcsname\relax
  \def\bibnamefont#1{#1}\fi
\expandafter\ifx\csname bibfnamefont\endcsname\relax
  \def\bibfnamefont#1{#1}\fi
\expandafter\ifx\csname citenamefont\endcsname\relax
  \def\citenamefont#1{#1}\fi
\expandafter\ifx\csname url\endcsname\relax
  \def\url#1{\texttt{#1}}\fi
\expandafter\ifx\csname urlprefix\endcsname\relax\def\urlprefix{URL }\fi
\providecommand{\bibinfo}[2]{#2}
\providecommand{\eprint}[2][]{\url{#2}}

\bibitem{bond94} Bond, J. R., Crittenden, R., Davis, R. L., Efstathiou, G.,  Steinhardt, P. J. (1994), \textit{Phys. Rev. Lett.}, \textbf{72}, 13
\bibitem{be84} Bond, J. R., Efstathiou, G. (1984) \textit{ApJLett},\textbf{285}, L45 
\bibitem{cash79} Cash, W. (1979)  \textit{ApJ}   \textbf{228} 939--947
\bibitem{christensenmeyer} {{Christensen}, N. and {Meyer}, R.} (2001) \textit{PRD}, \textbf{64}, 022001
\bibitem{cosmomc} \verb1http://cosmologist.info/cosmomc/1
\bibitem{cosmomcpaper} Lewis, A., Bridle, S. (2002) \textit{Phys. Rev. D}, \textbf{66}, 103511
\bibitem{healpixpaper}{{G{\'o}rski}, K.~M. and {Hivon}, E. and {Banday}, A.~J. and 
	{Wandelt}, B.~D. and {Hansen}, F.~K. and {Reinecke}, M. and 
	{Bartelmann}, M.}(2005) \textit{ApJ} \textbf{622} 759-771
\bibitem{converge} Gelman A., Rubin D. (1992) \textit{Statistical Science}, \textbf{7}, 457
\bibitem{Fisher} Fisher R.A. (1935) \textit{J. Roy. Stat. Soc.}, \textbf{98}, 39 
\bibitem{gilk} Gilks, W. R., Richardson, S., \& Spiegelhalter, D. J. (1996), Markov Chain Monte Carlo in Practice (London:  Chapman and Hall) 

\bibitem{hmc1} Hajian A. (2007) \textit{Phys. Rev. D}, \textbf{75}, 083525 
\bibitem{hmc2}  Taylor J. F.,  Ashdown M. A. J., Hobson M. P. (2007) arXiv:0708.2989
\bibitem{hstkey} Freedman et al. (2000) \textit{ApJ}, \textbf{553}, 47--72
\bibitem{healpixweb} \verb1http://healpix.jpl.nasa.gov/1
\bibitem{hivon02}{{Hivon}, E. and {G{\'o}rski}, K.~M. and {Netterfield}, C.~B. and 
	{Crill}, B.~P. and {Prunet}, S. and {Hansen}, F.} (2002) \textit{ApJ}   \textbf{567}  2--17
\bibitem{cmbwarp} Jimenez R., Verde, L. , Peiris H., Kosowsky A.  (2003) \textit{PRD}, \textbf{70}, 3005 
\bibitem{kendallstewardt77} {{Kendall}, M. and {Stuart}, A.}, "{The advanced theory of statistics."}, {London: Griffin, 1977, 4th ed.}

\bibitem{knoxthesis} {{Knox}, L.}(1995) \textit{PrD}, \textbf{52} , {4307-4318}.

\bibitem{kmj02} Kosowsky, A., Milosavljevic, M., Jimenez, R. (2002)  \textit{Phys. Rev. D}, \textbf{66}, 63007 

\bibitem{LAMBDA}  \verb1http://lambda.gsfc.nasa.gov/1
\bibitem{MS02} Martinez V. and Saar E. (2002), ''Statistics of the
Galaxy Distribution"  \textit{Chapman \& Hall/CRC Press}

\bibitem{page07} {{Page}, L. and {Hinshaw}, G. and {Komatsu}, E. and {Nolta}, M.~R. and 
	{Spergel}, D.~N. and {Bennett}, C.~L. and {Barnes}, C. and {Bean}, R. and 
	{Dor{\'e}}, O. and {Dunkley}, J. and {Halpern}, M. and {Hill}, R.~S. and 
	{Jarosik}, N. and {Kogut}, A. and {Limon}, M. and {Meyer}, S.~S. and 
	{Odegard}, N. and {Peiris}, H.~V. and {Tucker}, G.~S. and {Verde}, L. and 
	{Weiland}, J.~L. and {Wollack}, E. and {Wright}, E.~L.} (2007) \textit{ApJS},  \textbf{170} 335--376

\bibitem{Peebles80} {{Peebles}, P.~J.~E.}, "{The large-scale structure of the universe}", Princeton, N.J., Princeton University Press, 1980.

\bibitem{numrec} {{Press}, W.~H. and {Teukolsky}, S.~A. and {Vetterling}, W.~T. and 
	{Flannery}, B.~P.}, "Numerical recipes in FORTRAN. The art of scientific computing Cambridge: University Press, 1992.
	
\bibitem{spergel03} {{Spergel}, D.~N. and {Verde}, L. and {Peiris}, H.~V. and {Komatsu}, E. and 
	{Nolta}, M.~R. and {Bennett}, C.~L. and {Halpern}, M. and {Hinshaw}, G. and 
	{Jarosik}, N. and {Kogut}, A. and {Limon}, M. and {Meyer}, S.~S. and 
	{Page}, L. and {Tucker}, G.~S. and {Weiland}, J.~L. and {Wollack}, E. and 
	{Wright}, E.~L.}  (1979) \textit{ApJS}   \textbf{148}  175--194

\bibitem{spergel07} {{Spergel}, D.~N. and {Bean}, R. and {Dor{\'e}}, O. and {Nolta}, M.~R. and 
	{Bennett}, C.~L. and {Dunkley}, J. and {Hinshaw}, G. and {Jarosik}, N. and 
	{Komatsu}, E. and {Page}, L. and {Peiris}, H.~V. and {Verde}, L. and 
	{Halpern}, M. and {Hill}, R.~S. and {Kogut}, A. and {Limon}, M. and 
	{Meyer}, S.~S. and {Odegard}, N. and {Tucker}, G.~S. and {Weiland}, J.~L. and 
	{Wollack}, E. and {Wright}, E.~L.}, (2007)
                  \textit{ApJS} \textbf{170}, 377--408
                  
                  
\bibitem{verde03} {{Verde}, L. and {Peiris}, H.~V. and {Spergel}, D.~N. and {Nolta}, M.~R. and 
	{Bennett}, C.~L. and {Halpern}, M. and {Hinshaw}, G. and {Jarosik}, N. and 
	{Kogut}, A. and {Limon}, M. and {Meyer}, S.~S. and {Page}, L. and 
	{Tucker}, G.~S. and {Wollack}, E. and {Wright}, E.~L.},  (2003)   \textit{ApJS} \textbf{148}, 195--211

\bibitem{statastro} Wall J. V., Jenkins C. R. (2003) ''Practical statistic s for Astronomers", Cambridge University Press.	               

\end{thebibliography}

\newpage

\begin{table}
\newpage
\caption{$\Delta \chi^2$ for  the conventionals $1,2$,and $3-\sigma$ as a function of the number of parameters for the joint confidence levels.}
\label{table:chisq}
\begin{tabular}{|c|c|c|c|c|}
\hline
$\sigma$&p&1&2&3\\
\hline
1-$\sigma$&68.3\%&1.00&2.30&3.53\\
                     & 90\%& 2.71&4.61 & 6.25\\
2-$\sigma$&95.4\%&4.00&6.17&8.02\\
3-$\sigma$&99.73\%&9.00&11.8&14.2\\
\hline
\end{tabular}
\end{table}

\end{document}